\documentclass[sigconf,balance=false,prologue,table]{acmart}
 \usepackage{popets}
\setcopyright{popets}
\copyrightyear{YYYY}

\acmYear{YYYY}
\acmVolume{YYYY}
\acmNumber{X}
\acmDOI{XXXXXXX.XXXXXXX}
\acmISBN{}
\acmConference{}
\usepackage{pdflscape}
\usepackage{multirow}
\usepackage{array}
\usepackage{booktabs}
\usepackage{tabularx}
\usepackage{caption}
\usepackage[toc,page]{appendix}
\usepackage{longtable}
\usepackage{tikz}
\usepackage{amsmath}
\usepackage{graphicx}
\graphicspath{{images/}}
\DeclareGraphicsExtensions{.pdf,.jpg,.png}
\usepackage{wrapfig}
\usepackage[font=small,skip=2pt]{caption}
\usepackage{subcaption}
\usepackage{cleveref} 
\usepackage{float}
\usepackage{url}

\usepackage{color,soul}
\usepackage{todonotes}

\newcommand{\pp}[1]{{\scriptsize\emph{\sbox0 {\color{blue}Primal: #1}}}}
\usepackage{xspace}
\newcommand{\system}{PrivacyLens\xspace}

\newcommand{\hide}[1]{} 

\author{
    Aamir Hamid\textsuperscript{1,*},
    Hemanth Reddy Samidi\textsuperscript{1},
    Tim Finin\textsuperscript{1}, 
    Primal Pappachan\textsuperscript{2,†},
    Roberto Yus\textsuperscript{1}
    \\
    \textsuperscript{1}University of Maryland, Baltimore County, \textsuperscript{2}Portland State University
    \\
    \{ahamid2, finin, hsamidi1, ryus\}@umbc.edu, primal@pdx.edu
}

\begin{document}

\title{A Study of the Landscape of Privacy Policies of Smart Devices}

\maketitle


\section{abstract}
As the adoption of smart devices continues to permeate all aspects of our lives, user privacy concerns have become more pertinent than ever. Privacy policies outline the data handling practices of these devices. Prior work in the domains of websites and mobile apps has shown that privacy policies are rarely read and understood by users. In these domains, automatic analysis of privacy policies has been shown to help give users appropriate insights. However, there is a lack of such an analysis in the domain of smart device privacy policies. This paper presents a comprehensive study of the landscape of privacy policies of smart devices. We introduce a methodology that addresses the unique challenges of smart devices, by finding information about them, their manufacturers, and their privacy policies on the Web. Our methodology utilizes state-of-the-art analysis techniques to assess readability and privacy of smart device policies and compares it policies of e-commerce websites and mobile applications. Overall, we analyzed 4,556 smart devices, 2,211 manufacturers, and 819 privacy policies. Despite smart devices having access to more intrusive data about their users (using sensors such as cameras and microphones), more than 1,167 of the analyzed manufacturers did not have policies available. The study highlights that significant improvement is required on communicating the data management practices of smart devices.


\keywords{IoT, Privacy Policies, Data Protection Regulations}

\section{Introduction}
The Internet of Things (IoT) has gained rapid popularity in recent years. \textit{Smart} IoT devices are now utilized in transportation, industrial processes, smart homes, and health care. Smart devices have advanced capabilities that include, in general, the collection of information, usage of cutting-edge technologies, such as Artificial Intelligence (AI) to process such data, and automation of tasks to provide personalized user experiences~\cite{alter2016smart}. More than 40 million households in the US have adopted smart home devices, which is expected to reach 64.1 million by 2025~\cite{oberlo_smart_home_statistics}. 
The growing use of smart technology also poses privacy risks~\cite{apthorpe2017always,weber2010internet}. IoT devices collect large amounts of diverse data, and consumers sometimes do not understand what data is being collected in their environment~\cite{gubbi2013internet}. Data collection from IoT devices could lead to unbounded profiling of customers by businesses or disruption of regular operations by malicious entities~\cite{biocco2018study, atzori2010internet}. Consequently, consumers are concerned about the privacy risks of owning and using IoT devices~\cite {acquisti2015privacy}.

Privacy policies have traditionally provided information on the data collected/used/shared by services such as e-commerce and mobile applications, and extensive analysis has been performed on them~\cite{fabian2017large,mcdonald2009score,krumay2020readability}. However, to the authors' knowledge, privacy policies of IoT devices (e.g., smart home devices) have not received as much attention despite the privacy challenge they pose. Arguably, the most important work is that of the Mozilla Privacy Not Included project~\cite{mozilla_privacy_not_included} in which human analysts study smart devices w.r.t their privacy, security, and usage of AI. However, manual analysis is not scalable to the increasing number of IoT devices in the market. Mozilla reported that their human analysts spent 68,160 minutes (47 days) reading and analyzing privacy policies in 2022. Hence, automatic collection and analysis of IoT privacy policies is required. 
While state-of-the-art analysis techniques for privacy policies used in other domains could be used for smart device policies, they present a unique challenge in finding and collecting  them.
In other domains where privacy policies have been studied, such as websites and mobile applications, the relevant policies are associated with the website or application of interest, making it relatively easy to find and collect those policies.
For smart devices, this presents a unique challenge because while the devices are sold on e-commerce websites, their privacy policies resides on the manufacturer website.
Thus finding and collecting policies in the smart devices domain requires an entirely new approach.

This study aims to bridge the existing knowledge gap by providing an in-depth analysis of smart device privacy policies. We are specifically interested in understanding how easy it is for users to find these policies at the time of purchasing a smart device, how difficult these policies are to be processed, what these policies describe about the data collection practices, and how these policies compare with established domains such as mobile apps and Web services. Accordingly, our study aims to answer two main research questions: \textbf{What is the state of smart device privacy policies?} and \textbf{How do smart device privacy policies compare to those of more consolidated domains?}
To answer the questions, we integrate smart device privacy policy collection techniques with existing methodologies for analysis to derive insights. In particular, we develop and implement a framework, \system, based on state-of-the-art techniques, which automates the collection, analysis, and publication of insights about smart device privacy policies. \system searches e-commerce websites for smart devices, extracts metadata, finds their manufacturers' websites, and retrieves privacy policies from them. It uses the Wayback Machine~\cite{waybackmachine} to obtain archived past privacy policies for the manufacturers to enable longitudinal analysis, such as policy evolution and the impact of privacy regulations. Using natural language processing (NLP) and machine learning (ML) techniques, \system extracts different features of each privacy policy, including their overall quality, readability, and ambiguity. 
In summary, the main contributions of this paper are as follows:

\begin{itemize}
    \item This is, up to the author's knowledge, the first comprehensive analysis of the landscape of privacy policies of smart devices involving more than 4,556 smart devices from 28 categories, 2,211 manufacturers, and 1,131 privacy policies (from which 312 are archived versions).
    \item We compare smart device policies against 100 privacy policies of the major e-commerce websites and 350 privacy policies of current Android applications.
    \item We introduce a methodology to find, extract, and analyze smart device privacy policies, including past versions, using state-of-the-art techniques.
    \item We develop an open source framework implementing this methodology that can be used for future analysis.
\end{itemize}

Additionally, we provide a discussion with several important observations from the study focusing on the challenges that customers might face finding and understanding the privacy policies of smart devices. We also discuss about the observed similarities with other domains which indicate the usage of policy 'templates'.

The rest of the paper is structured as follows. 
\Cref{sect:relatedwork} reviews the state of the art in privacy policy collection and analysis.
\Cref{sect:methodology} describes the methodology used to conduct the study.
\Cref{sect:analysis1} and \Cref{sect:analysis2} explain findings and insights from privacy policies. In Section \Cref{sect:discussion}, we provide a discussion on challenges and opportunities. Finally, Section \Cref{sect:conclusion} concludes the paper and presents directions for future research

\section{Related Work}
\label{sect:relatedwork}
Privacy policy research increasingly relies on automated frameworks and content analysis to thoroughly assess policy clarity, compliance, data practices, and user-centric aspects across different sectors. While e-commerce and mobile applications have been extensively studied, the smart device domain remains less explored. The notable exception is the work of Kuznetsov et al.~\cite{kuznetsov2022privacy}, who analyzed 592 privacy policies from various IoT device manufacturers. Our approach, similar to theirs, utilizes e-commerce websites to identify smart devices and extract relevant manufacturer privacy policies. However, our collection framework includes unique features: a module to determine if a device qualifies as "smart" (which resulted in a larger number of products analyzed), language detection to ensure policy texts are in English, and automatic segmentation of privacy texts. Diverging from Kuznetsov et al., we also collect and analyze historical versions of these policies to trace their evolution. Our analysis not only examines the readability of the privacy policies but also delves into their data management practices and updates, offering a comprehensive overview of the privacy landscape in the smart device sector.

As mentioned before, e-commerce and websites have received special attention. For instance, in~\cite{AIMEUR2016368} the authors examined the impact of privacy policy formats on user trust in online services. Their study involved 717 participants and compared conventional policies against those offering greater user control and personalization. In~\cite{8049146}, the authors analyzed privacy policies from e-commerce websites to understand data collection and use purposes. Employing content analysis, they identified six unique data purpose categories and observed specific language patterns for expressing these. 
In~\cite{krumay2020readability} the authors investigated various quantitative approaches to measure privacy policy readability. It identified the challenge of varying results based on measurement methods and proposed a combined system for a comprehensive company assessment. 
In \cite{sunyaev2015availability}, the authors investigated the prevalence and quality of privacy policies for mobile health apps on iOS and Android platforms. They reviewed 600 apps to assess the availability, scope, and transparency of their privacy policies. 
\cite{Bui2021AutomatedEA} investigated the impact of privacy policies on users' understanding of privacy practices. The authors analyzed the effects of data practice annotation, highlighting and describing the extracted data practices to help users better understand privacy-policy documents.

The approach in~\cite{Sadeh2014TowardsUP} enhances previous research in NLP, privacy preference modelling, crowdsourcing, and privacy interface design, specifically targeting website privacy policies. It extends existing research on user preference modelling in website privacy policies and incorporates innovative approaches such as semi-automated feature extraction and privacy notice design based on extracted policy features. Polisis~\cite{10.5555/3277203.3277243} uses a neural network hierarchy to extract high-level privacy practices and precise data from 130k website privacy policies and has an interface for both structured and free-form querying of privacy policies. PI-Extract~\cite{Bui2021AutomatedEA} is a fully automated system to extract fine-grained personal data phrases and their corresponding practices from website privacy policies. It is based on a neural model that outperforms rule-based baselines in accurately extracting privacy practices. In study\cite{ibdah2021should}, the authors investigated user attitudes and perceptions toward privacy policies to enhance user awareness and understanding. They analyzed data from 655 participants to understand the factors influencing users' willingness to read privacy policies, including the effects of prior experiences like cyber-attacks and data-sharing practices. In~\cite{gebauer2023human}, the authors deal with the problem of extracting transparency information from website privacy policies by proposing a `Human-in-the-Loop' approach that combines machine learning-generated suggestions with human annotation decisions. Their prototype system streamlines the annotation process by providing meaningful predictions to users, resulting in improved performance compared to other extraction models for legal documents. In~\cite{290991}, the authors propose a method for analyzing privacy policies through an integrated approach. They introduce PoliGraph, a knowledge graph for mapping relations within privacy policy texts, and PoliGraph-er, an NLP tool for automatic PoliGraph generation. This study redefines ontologies to understand the context of privacy policies and application domains.

As we will describe in the following sections, we leverage and adapt some of the previously reviewed state-of-the-art techniques and combine them into a single framework to collect and analyze smart device privacy policies.



\section{Methodology}
\label{sect:methodology}

\begin{figure*}[ht]
\centering
\includegraphics[width=1\linewidth]{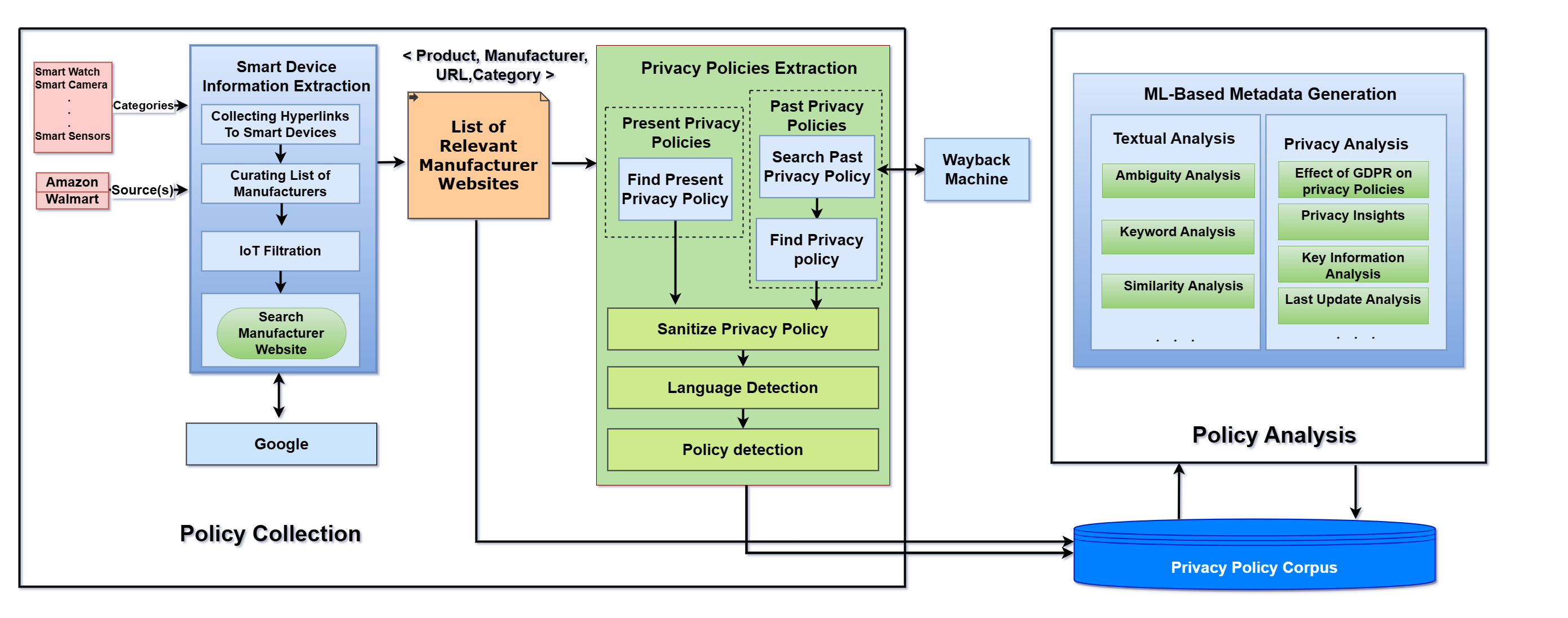}
\caption{A high-level overview of our methodology to find, collect, and analyze smart device privacy policies. }
\label{fig:architecture}
\end{figure*}

Our study's methodology (illustrated in Figure~\ref{fig:architecture}) follows the conventional two-step process of policy collection and analysis, as established in prior research~\cite{8049146,krumay2020readability,sunyaev2015availability}. We have developed a framework, \system, encapsulating all these steps. This framework is publicly available to support further privacy policy studies\footnote{\url{https://anonymous.4open.science/r/privacy-lens-81F4}}. Besides analyzing smart device privacy policies, we also use this tool to examine 350 Android app privacy policies (from~\cite{Zimmeck2019MAPSSP}) and 100 policies from top e-commerce websites listed on \textit{semrush}\footnote{\url{https://www.semrush.com/blog/most-visited-websites/}}. We also expanded our dataset by incorporating policies from 10 smart connected cars, recognizing them as integral components of the broader smart device ecosystem. These additional analyses facilitate a comparative study.


\subsection{Policy Collection}

\paragraph{IoT product information extraction.}
We designed and implemented a multi-threaded web scraper with a specialization in extracting data related to smart devices from popular e-commerce platforms such as Amazon and Walmart. The web scraper's code establishes a connection, utilizing \textit{WebDriverManager}\footnote{\url{https://bonigarcia.dev/webdrivermanager/}}, to a Firefox browser for data extraction. Subsequently, it uses \textit{Selenium}\footnote{\url{https://www.selenium.dev}} to manipulate the web browser programmatically, initiating searches for smart devices based on a predefined list of relevant categories as search queries. This list incorporates keywords prefixed with the terms 'Smart', 'Connected', 'WiFi-Enabled', 'Internet-Ready', 'IoT', 'Wireless', 'Cloud-Based', 'Bluetooth-Enabled', 'Remote-Controlled', 'Networked', 'Digital'. These prefixes are combined with various product types, including body scanners, cameras, connected vehicles, doorbells, entertainment devices, fitness equipment, gaming technology, health trackers, home devices, lights, location trackers, locks, monitors, mounts, networking devices, projectors, scales, security systems, sensors, speakers, thermostats, TVs, and watches, resulting in a total of 53 search keywords.

After executing each search, the system parses the resulting HTML code using \textit{BeautifulSoup}\footnote{\url{https://www.crummy.com/software/BeautifulSoup/}} to extract the URLs of individual products listed on the web page. We explore the first 40 pages of results per keyword. The system then navigates to each product's URL using the web driver to collect metadata, including the manufacturer's name and country of origin, from the HTML markup. Furthermore, the system extracts additional information, when available, to determine whether a specific product qualifies as a "smart" device. This determination is essential because e-commerce platforms sometimes return products that lack Internet connectivity and do not meet the criteria for being classified as "smart". For example, on Amazon, we aim to extract information contained in the description fields related to \textit{connectivity technology}, \textit{connectivity protocol}, and \textit{wireless communication technology}. Additionally, the system leverages textual descriptions of the products through preprocessing steps such as lemmatization, tokenization, removal of stopwords, and TF-IDF vectorization. Each product is assigned a score based on the frequency and importance of IoT-relevant keywords in its description, including ``wifi'', ``bluetooth'', ``voice-controlled'', and ``app-controlled'' (the complete list is available in Appendix Table~\ref{tab:iot_keywords}). To classify a product as a smart device, a threshold score of 0.4, determined experimentally, is employed. This threshold balances precision and comprehensive coverage in the identification process. 

At the conclusion of this stage, the system provides, for each identified smart device, its respective e-commerce URL, name, manufacturer, and country of origin. The initial set of 53 search keywords underwent a meticulous analysis and condensation process, resulting in the creation of 28 distinct categories, each dedicated to a specific type of device.

\paragraph{Manufacturer's website extraction.}
To locate the manufacturer's website, \system executes a Google search query that combines the manufacturer's name and device type (e.g., 'fitbit' + 'smart watch'). Next, the web scraper analyzes the HTML markup of the first page of results to identify all URLs. For each URL, \system calculates a value representing the likelihood of it being the official manufacturer's website. This determination is made by comparing the manufacturer's name to the domain using the Ratcliff/Obershelp algorithm~\cite{ratcliff1988pattern}. The URL with the highest calculated value is selected if it surpasses a predefined threshold (0.8), which we determined experimentally (see Appendix~\ref{appendix:a} for further details).

\paragraph{Privacy policy extraction and cleaning.}
Once the website's URL is obtained, our system proceeds to parse its HTML content using BeautifulSoup. The goal is to identify "href" elements that are likely to point to privacy policies. To achieve this, the system employs a predefined set of keywords, which includes terms like "privacy", "policy", "data", and "protection". This keyword search is further enriched with multilingual options to broaden the scope of the search. The script analyzes both hyperlink texts and URLs using a two-step approach, English and Multilingual, alongside a contextual analysis that assesses the adjacent text and HTML structure to determine link relevance to privacy policies. Following this analysis, a validation and normalization process is applied to correct any relative links. This ensures that all links are appended to the site's base URL, guaranteeing the accurate retrieval of complete and functional privacy policy URLs.

The text extraction task is carried out using boilerpy3\footnote{\url{https://pypi.org/project/boilerpy3/}}, a version of the Boilerpipe library, and it employs the Canola extractor\footnote{\url{https://rdrr.io/rforge/boilerpipeR/man/CanolaExtractor.html}} due to its efficiency in isolating relevant content from web pages. The process begins with a GET request to each URL, and the module handles issues such as malformed URLs or incorrect HTTP responses adeptly. The Canola extractor then analyzes the web page, effectively separating the main content from extraneous elements such as ads, navigation links, and headers. It utilizes algorithms that assess text density, page layout, and HTML indicators to precisely identify the core text. This identified text is processed using the Canola extractor's \verb|"get_content"| function, which is essential for detecting and rectifying any extraction errors. The resulting output is a clean and accurate extraction of the text content, prepared for further analysis.

\paragraph{Language detection.}
The objective of language detection was to ensure that we exclusively review privacy policies that are available in English.
Our language detection method employs a majority voting system combining six libraries:
Langdetect\footnote{\url{https://pypi.org/project/langdetect/}}, CLD3\footnote{\url{https://pypi.org/project/pycld3/}}, LangID\footnote{\url{https://pypi.org/project/langid/}},
Guess-language\footnote{\url{https://pypi.org/project/guess-language/}}, fastText\footnote{\url{https://pypi.org/project/fasttext/}}, and Textacy\footnote{\url{https://pypi.org/project/textacy/}}. Each has a unique approach: Langdetect uses n-gram models and probabilistic algorithms; CLD3 excels in web content with compact algorithms; LangID and Guess-language offer quick identification with heuristic methods; FastText's deep learning is effective for context-rich text; Textacy analyzes linguistic features. The system aggregates each library's independent language predictions, determining the final language based on the most common prediction.

\paragraph{Policy detection.}
We employ the method outlined in \cite{inproceedings_2} for policy detection, focusing on key-phrase extraction and efficient feature selection. This approach uses a uniform threshold for key phrase extraction across different algorithms, complemented by comprehensive lemmatization to minimize feature redundancy. Adopting a binary feature matrix, which marks the presence or absence of key phrases, avoids the limitations of frequency-based methods. Feature selection through ANOVA F-value is instrumental in isolating the most relevant phrases. Additionally, an ensemble soft voting classifier, combining multiple models like the linear support vector machine, random forest, and logistic regression, enhances accuracy and adaptability in diverse linguistic contexts. Our application of these techniques aims to achieve a more precise identification of privacy and cookie policies.

\paragraph{Past privacy policy extraction.} To gather historical privacy policies, \system uses the Wayback Machine \cite{waybackmachine}, an Internet archive containing older snapshots of websites. For each manufacturer URL obtained in the previous step, \system queries the Wayback Machine to get snapshots at different points in time\footnote{We opt to use the manufacturer's URL rather than the privacy policy URL since we observed that privacy policy URLs changed over the years on a significant number of sites.}. \system initiates one query per year for the past ten years, in addition to additional queries for the months leading up to and following the implementation of significant data privacy regulations, such as the GDPR~\cite{voigt2017eu} and CCPA~\cite{ccpa}. Once the different snapshots are retrieved, \system uses the process explained before to locate the privacy policy within the website and then clean it for further analysis.

\pp{After reading this subsection, I am still uncertain whether you analyze the privacy policy of the manufacturer, device, or both? For e.g., Google/Fitbit or Alexa/Amazon. To make this clear, do you mind adding examples in brackets to each of the steps as you have done in extraction and IoT filtration steps?}

\subsection{Policy Analysis}
Our methodology employs Natural Language Processing (NLP) and Machine Learning (ML) techniques to analyze privacy policies, extracting key insights. The objective is to assess the document's readability and evaluate its compliance with privacy standards.

\subsubsection{Privacy Analysis}
We extract the following privacy features of a smart device policy.

\paragraph{Topic Coverage and Keyword Usage.}
We employed Natural Language Processing (NLP) techniques, including stop words removal and lemmatization, to preprocess privacy policy texts.
Our methodology integrated seed keywords, drawing on previous research~\cite{wilson2016creation,inproceedings_1}, and added two new categories focusing on IoT-related data practices~\cite{9581254} and legal compliance (as detailed in~\Cref{tab:privacy_keywords}). 
We used Latent Dirichlet Allocation (LDA) for topic modeling to uncover thematic structures within the privacy policies, enabling comparisons of keyword coverage across datasets.
For quantitative analysis, sklearn's CountVectorizer tool\footnote{\url{https://scikit-learn.org/stable/modules/generated/sklearn.feature_extraction.text.CountVectorizer.html}} measured the frequency of our extended seed keywords. This facilitated a systematic comparison of keyword prevalence and thematic emphasis across different privacy policy categories, providing a comprehensive understanding of their content focus.

\paragraph{Privacy Policy Similarity and Clustering}
Our methodology uses a mixed-methods approach to analyze and cluster privacy policies from diverse digital platforms, including smart devices, mobile apps, and e-commerce websites. The approach combines semantic text similarity with quantitative clustering techniques. 
For the semantic text similarity analysis, we encoded privacy policy texts using the SentenceTransformer model \textit{all-MiniLM-L6v2}\footnote{\url{https://huggingface.co/sentence-transformers/all-MiniLM-L12-v2}} \pp{this reads more like "textual analysis" than "privacy analysis"?}. This model is particularly adept at mapping sentences and paragraphs to a dense vector space, facilitating a detailed analysis of semantic similarities. The choice of this model was driven by its advanced capabilities in semantic understanding, essential for accurately interpreting the complex language often found in privacy policies. The resulting tensor-encoded texts were then analyzed using a cosine similarity metric, with the findings visualized in a heatmap. This visualization clearly represented the semantic relationships between different policies, aiding in identifying common patterns and structures.

In the second phase, we concentrated on clustering the privacy policies. This involved categorizing the policies and processing them for textual analysis, followed by using sklearn's CountVectorizer to filter out stopwords, ensuring a clean dataset for subsequent steps, and adjusting the data to align with the corpus' structure, which revealed around 7,000 distinct features per policy.  We employed UMAP\cite{McInnes2018UMAPUM}, known for its efficiency in handling complex data, to transform the texts for clustering analysis. We choose HDBSCAN\cite{McInnes2017hdbscanHD} for its ability to handle complex, hierarchical data structures in an unsupervised clustering approach. Our analysis utilized two clustering methods: one based on the textual content of the policies and the other incorporating additional extracted features like reading level and entropy. This dual approach enabled us to compare the effectiveness of text-based clustering against a feature-enhanced clustering method, providing a comprehensive understanding of the underlying groupings in privacy policy texts.

\paragraph{Tracking Changes in Privacy Policies}
Regular updates to privacy policies are crucial for keeping users informed about the handling and use of their data. Our method uses advanced NLP techniques, including context analysis and pattern matching, to pinpoint the exact dates of policy updates and monitor any changes, offering a detailed overview of these modifications.
The introduction of new data protection laws is a key factor influencing policy changes. Presently, over 150 countries have enacted such regulations~\cite{banisar2023national}. \system leverages a deep learning approach with the Bidirectional Encoder Representations from Transformers (BERT) model~\cite{bert2019}, fine-tuned using the Semantic Textual Similarity Benchmark dataset (STS-B)~\cite{cer-etal-2017-semeval}. This method assesses alterations in smart device privacy policies before and after the implementation of specific regulations (e.g., GDPR). By computing a semantic similarity score for policy pairs (pre- and post-event), we quantitatively measure the extent of these changes.

\subsubsection{Readability Analysis}

\system analyzes policy readability extracting eight features.

\paragraph{Entropy.} In privacy policy analysis, entropy is a vital metric quantifying textual uncertainty and complexity, highlighting potential interpretive challenges for users. It represents the average information produced per letter in a text, reflecting the uncertainty or disorder within a document. We use Shannon's method to calculate the entropy of English text~\cite{shannon1951prediction}.

\paragraph{Reading Time.} Privacy policies, often lengthy and time-consuming, challenge users in making informed decisions~\cite{mcdonald2008cost}. \system calculates reading time using an average pace of 238 words per minute (WPM)~\cite{brysbaert2019many}.

\paragraph{Unique Words.} 
Privacy policies often utilize technical jargon to explain data usage and control, making them difficult for consumers to understand. Unique, low-frequency words in these documents are key to understanding, as they provide essential context and learning aspects~\cite{hiebertunique}. However, their rarity makes comprehension challenging, as it requires a deep understanding of specialized vocabulary.
To analyze these policies, the text is first converted to lowercase, with stop words, punctuation, and numbers removed. It is then tokenized~\cite{geitgey2018natural} for standardization using the Spacy library~\cite{spacy2}. Afterwards, \system identifies unique words -- those with distinct character sequences -- and counts them to determine the document's distinct vocabulary size. Additionally, the system calculates the ratio of unique word count to total word count, providing a quantitative measure of the text's lexical diversity.

\paragraph{Coherence Score.} 
Topic modeling uses a coherence score, to measure how comprehensible a topic is to people, based on word similarity within a topic and their document frequency~\cite{syed2017full}. The Latent Dirichlet Allocation (LDA) method, a form of unsupervised machine learning, assists in text analysis by determining the most representative topics~\cite{yu2001direct}. This technique creates a Dirichlet distribution for documents across topics, with topics and words emerging from multinomial distributions. The coherence score, summing up scores between all word pairs, assesses the quality of topics learned. The CV measure, which utilizes cosine similarity and normalized pointwise mutual information (NPMI) based on word co-occurrences, calculates this score. The overall coherence of a privacy policy is then computed as $\sum_{i < j} \text{score}(w_i, w_j)$, where $w_i$ and $w_j$ are words at positions i and j, respectively, in the text.
Regarding readability, a high coherence score implies a well-structured and clear flow of ideas, enhancing readability. Conversely, a low score might indicate a disjointed or unclear thought progression, making the text harder to comprehend.

\paragraph{Frequency of Imprecise Words.}
Vague language, using words like "commonly" or "normally," can lead to ambiguity in understanding a service provider's operations. 
\system utilizes NLTK for text tokenization and regex to count the frequency of such imprecise words (see \Cref{table2} for the full list), thereby quantifying their presence in a privacy policy. 

\paragraph{Connective Words Frequency.}
Although connective words like "and" or "then" aid in forming coherent sentences, their excessive use can lead to text complexity.
\system counts the frequency of such connective words, using a predefined list (see \Cref{conn}), similar to the approach used for the previous feature.

\paragraph{Grammatical Errors.}
The integrity of a work depends on proper grammar, much as it does on word spelling~\cite{naber2003rule}. \system takes privacy document as input and then uses the NLTK library for tokenization (i.e., breaking the text into sentences) and the \url{language_tool_python} library to check for grammatical errors. It counts the total number of sentences and the number of sentences that contain at least one mistake, providing a measure of the grammatical correctness of the input text.

\paragraph{Readability.} Readability signifies how easily a text, like a policy, can be understood based on its vocabulary, syntax, and sentence structure. Various readability tests exist~\cite{kouame2010using}, devised by linguists, each considering different text aspects. We employed the Flesch-Kincaid Grade Level \cite{flesch2007flesch}, which presents the score as a U.S. grade level. This metric represents the educational level required to understand a text and is computed using the following formula:
\begin{equation} \label{eq1}
\resizebox{0.9\hsize}{!}{$FKGL = 0.39 \left(\frac {total words}{total sentences}\right) + 11.8 \left(\frac {total syllables}{total words}\right) - 15.59$}
\end{equation}
\noindent Here 0.39 and 11.8 are weights for the average sentence length and syllables per word, respectively, and -15.59 calibrates the score to U.S. grade levels.

\paragraph{Ambiguity.} While privacy policies should clearly state the data handling practices, they often contain ambiguous language~\cite{reidenberg2016ambiguity}. \system uses a supervised learning approach to classify a privacy policy based on a scale with three ambiguity levels (\textit{not ambiguous}, \textit{somewhat ambiguous}, and \textit{very ambiguous}) presented in~\cite{kotal2021effect}. We annotated 100 IoT device policies extracted from \system and trained a random forest classifier~\cite{breiman2001random} and a Support Vector Classifier~\cite{cortes1995support} classifier. SVC works by finding the hyperplane that best separates different classes in the feature space, making it effective for complex classification problems. Random Forest is an ensemble learning method that builds multiple decision trees and merges them to get a more accurate and stable prediction, making it robust against overfitting and effective in handling large datasets with higher dimensionality. \system applies both classifiers to each privacy policy and stores their output labels.



\section{Analysis of Smart Device Policies}
\label{sect:analysis1}
This section analyzes the smart device privacy policies extracted and their historical evolution. The data used in this analysis can be accessed and visualized at \url{https://privacy-lens.web.app/home}.

\subsection{How Difficult is it to Find the Policies?}

Our methodology for privacy policy searching and collection (see Section~\ref{sect:methodology}) resulted in 4,556 smart devices (after filtering). Remember that this number of devices covers everything in the 40 first pages of results of the e-commerce platform for each of the 53 search queries run. Hence, we believe that this dataset contains the majority of smart devices in the market at the moment the study was performed. We split the smart devices into 28 categories, including smart cameras, smart home devices, and smart lights (see Appendix Table \ref{tab:cat_freq} for all the categories).

From the list of devices, we identified 2,211 unique manufacturers. The minimum, maximum, and median number of devices per manufacturer are 1, 237, and 1, respectively. 
Among these manufacturers, we found a website for 1,044 (47\%) and could only find links to privacy policies in 906 (40\%). Additionally, in our experiment, we could only access 819 of those links (37\% of the manufacturers). Some were inaccessible due to broken links, HTTP errors, or removed content. This highlights an important problem with smart devices and privacy: most manufacturers do not have an online privacy policy for their products that customers can access before a purchase. We acknowledge that, in some cases, the privacy policy might be available in some other form once the device is purchased. 

\begin{figure}[!htb]
    \centering
    \includegraphics[width=0.48\textwidth]{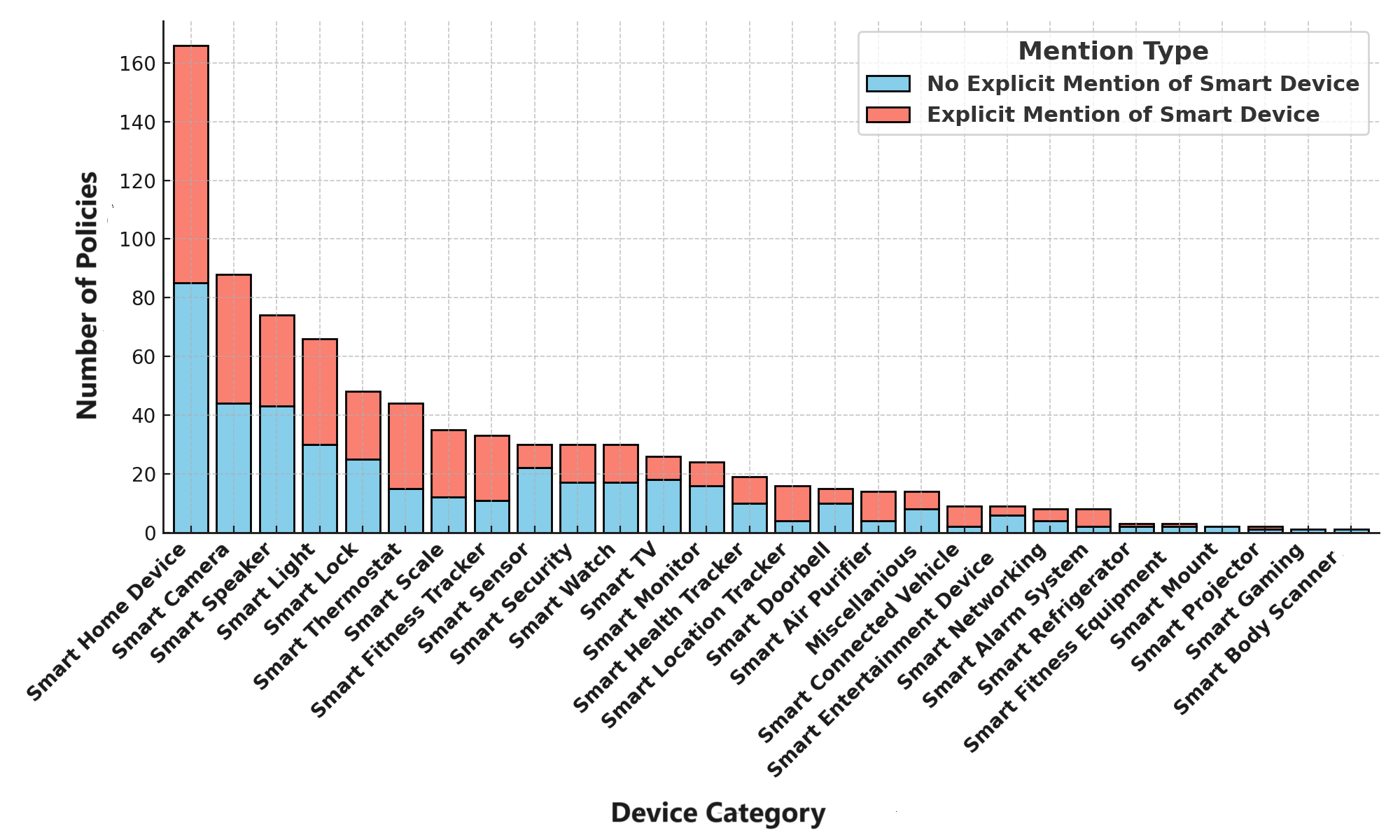}
    \caption{Distribution of policies based on the type of smart device.}
    \label{fig:my_device}
\end{figure}

Of 819 smart device privacy policies, only 405 explicitly mention that the policy applies to the smart device itself. This shows that most of the privacy policies either do not explicitly mention or just describe the data management practices of the website. 
Figure~\ref{fig:my_device} shows the number of privacy policies per smart device category and the number explicitly mentioning the device. ``Smart Home Device'' is the category with the largest number of products for which we could find a privacy policy followed by cameras and speakers. The distribution between explicit and non-explicit mentions of the device within the policy is balanced for the three. The categories with the highest percentage of policies with explicit mentions are ``Smart Light'' and ``Smart Thermostat'', which generally are devices without sensors that people might consider intrusive (like cameras or microphones). On the other hand, categories like "Smart Sensor" and "Smart Doorbell" have relatively few mentions. In general, we observe that many of the policies related to devices that people might consider intrusive (e.g., cameras, speakers, assistants, health trackers) do not explicitly mention the devices.

\begin{figure}[!htb]
    \centering    
    \includegraphics[width=0.5\textwidth,scale=1.5]{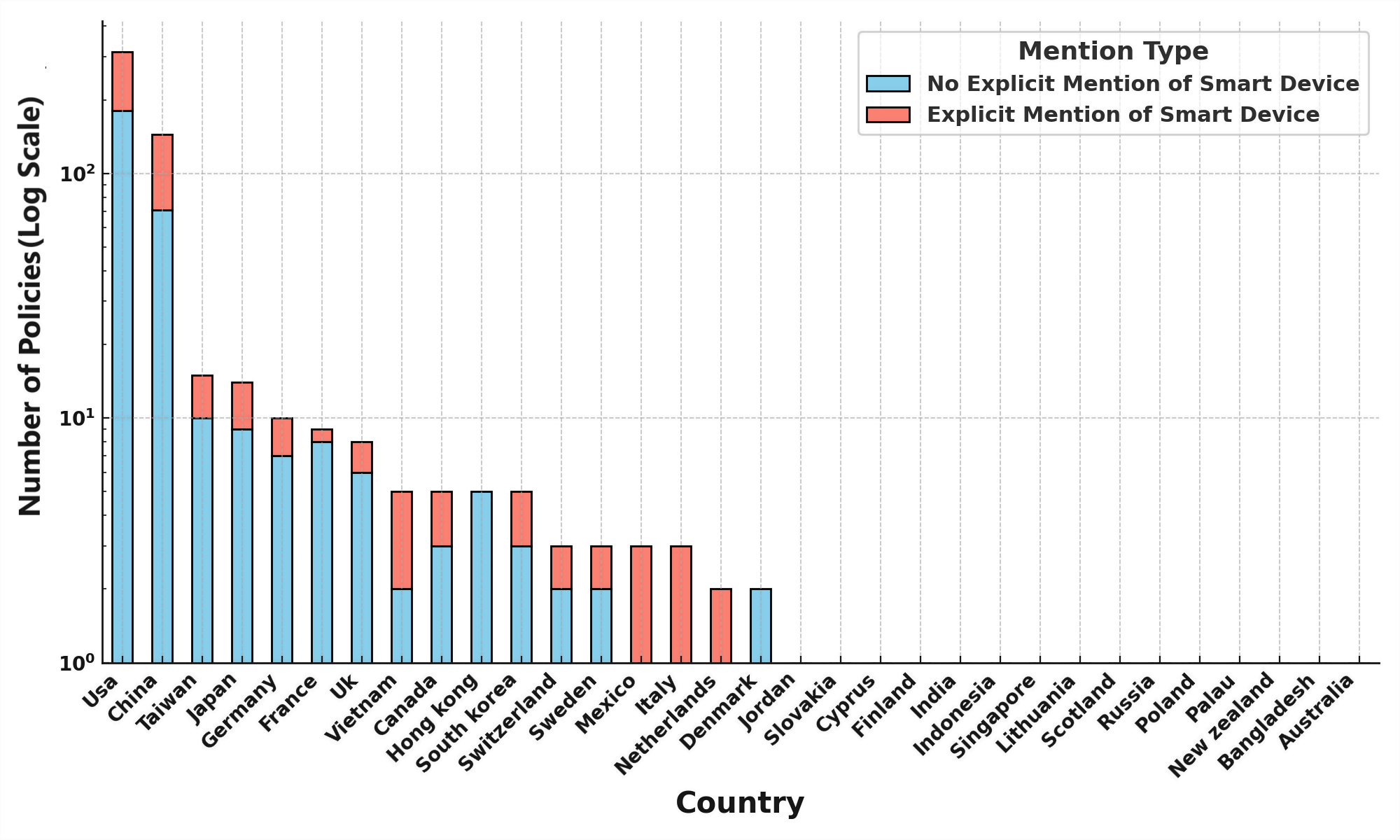}
    \caption{Distribution of policies based on manufacturer's country. }
    \label{fig:my_country}
\end{figure}

We further investigate the distribution of policies depending on their mention type by country of origin of the smart device manufacturer (see Figure \ref{fig:my_country}). Note that the figure's Y-axis is on a logarithmic scale due to some countries' small number of policies. We first observe that most manufacturers are from the USA (311) or China (144). The largest and lowest categories of devices manufactured by companies in the USA are ``Smart Home Devices'' (25.08\%) and ``Smart Lights'' (6.43\%). In the case of China, the categories are ``Smart Cameras'' (16.67\%) and ``Smart Lights'' (11.81\%). Most countries contained only one manufacturer in the extracted dataset (e.g., Australia, Cyprus, Finland, Indonesia, Jordan, Lithuania, Singapore, and Slovakia). The countries with the highest number of policies explicitly mentioning smart devices are China (50.6\%) and the Netherlands (100\%). Note that the countries with the lowest mentions within the European Union include France and the UK, which is unexpected due to the stringent privacy legislation in the EU (i.e., GDPR).

\begin{figure}[!htb]
    \centering
    \includegraphics[width=\linewidth]{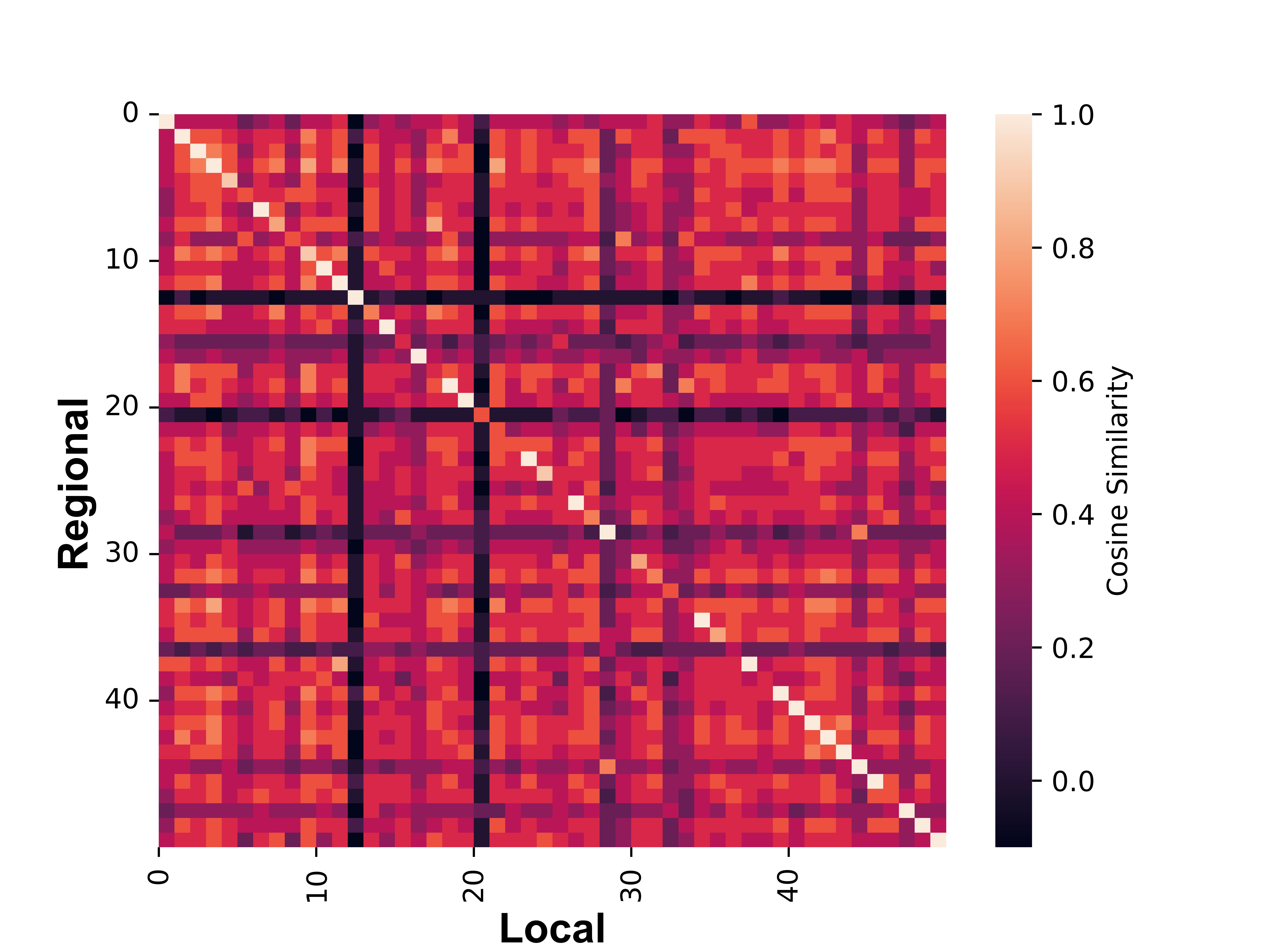}
    \caption{Heatmap of Regional Policy Variations for Smart Device Manufacturers.}
    \label{fig:regionlvpn}
\end{figure}

Since our analysis was executed in the US, we further explored whether the policies obtained would have been different from other regions. We randomly sampled 50 manufacturers and used VPNs to collect their policies simulating access from the country of origin of the manufacturer. We then used cosine distances to test for similarities. Our results, as shown in \Cref{fig:regionlvpn}, revealed that most smart devices (92\%) had consistent policies across regions, with some exceptions (8\%). In these cases, about 40\% of the policy content was similar, indicating changes were more about additions or deletions rather than complete rewrites. These variations seem to be strategic adjustments by manufacturers to comply with local laws and consumer preferences, while keeping a consistent core framework. For example, Hyundai's policies vary because the one showed in the US includes California-specific practices, in line with the state's privacy laws. Toyota's terms offer a broad legal framework, with additional details for the UK to comply with EU and UK data protection laws. AquaSound has a general privacy policy and a detailed cookie policy, likely due to EU GDPR requirements. Notably, most devices with regional policy differences are from Asia, reflecting manufacturers' efforts to meet both regional legal requirements and local data privacy concerns.

\subsection{How Readable are the Policies?}

Table\ref{table:fetaure_avg} contains a summary of the readability metrics extracted for the 819 smart device privacy policies. 
The Coherence Score, ranging from 0.25 to 0.73, assesses the logical flow and consistency of the document. The Frequency of Imprecise Words (0 to 0.04) and Connective Words (0.00 to 0.08) indicate the use of vague terms and logical connectors, respectively, which can impact clarity and understanding. Reading Complexity, ranging from 4.70 to 26.99, reflects the text's difficulty level, while Reading Time (1 to 130 minutes) estimates the time required for a complete read-through. Entropy, from 5.99 to 10.13,  measures the variability or unpredictability in word usage, suggesting complexity in terms of information content. The Frequency of Unique Words (0.09 to 0.67) indicates the diversity of vocabulary, and the presence of Grammatical Errors (0.05 to 1) is a direct measure of linguistic accuracy. When analyzing the ambiguity of the policies, we note that the number of policies classified as not ambiguous, somewhat ambiguous, and very ambiguous is 471, 186, and 162, respectively. In particular, when focusing on privacy policies explicitly mentioning the smart device, we observe that the distribution based on ambiguity is 242, 106, and 81, respectively.

In summary, on one hand, we observe that the average smart device privacy policy's text is not ambiguous, relatively short (it would take 11 minutes to be read), contains a very small number of imprecise words and grammatical errors, uses a minimal number of connective words (which translates into shorter sentences). On the other hand, we observe that the reading complexity is relatively high (a Flesch-Kincaid of 13 falls into the skilled level), and its coherence is relatively low. Considering an average reading speed of 240 words per minute for an 11-minute read, the document totals 2640 words. It has a 27\% uniqueness in its vocabulary, which consists of approximately 715 words. Notably, the 8-bit entropy per word is high, indicating a diverse and unpredictable vocabulary usage.

\begin{table}[!htb]
\centering
\resizebox{\linewidth}{!}{
\begin{tabular}{|l|l|l|l|}
\hline
\textbf{Policy Features} & \textbf{Min Value} & \textbf{Median Value} & \textbf{Max Value} \\
\hline
Coherence Score & 0.25 & 0.31 & 0.74 \\
\hline
Freq. of Imprecise Words & 0.00 & 0.02 & 0.04 \\

\hline
Freq. of Connective Words & 0.00 & 0.04 & 0.08 \\

\hline
Reading Complexity & 4.70 & 12.71 & 26.99 \\

\hline
Reading Time (Min) & 1.00 & 11.00 & 130.00 \\

\hline
Entropy & 5.99 & 8.03 & 10.13 \\

\hline
Freq. of Unique Words & 0.09 & 0.27 & 0.67 \\

\hline
Grammatical Errors & 0.0 & 0.007 & 0.01\\

\hline
\end{tabular}
}
\caption{Statistics for the readability analysis of the smart devices.}
\label{table:fetaure_avg}
\end{table}

In addition to the textual features extracted, further privacy-related information is detailed in Appendix Table \ref{tab:privacy_keywords}. This part of the extraction focuses on aspects of the privacy policies that directly pertain to how user data is handled, such as the types of data collected, data usage policies, sharing practices, and user choices.

\subsection{How Much Do the Policies Describe Data Management Practices?}

\begin{figure}[h!]
    \centering
    \includegraphics[width=\linewidth]{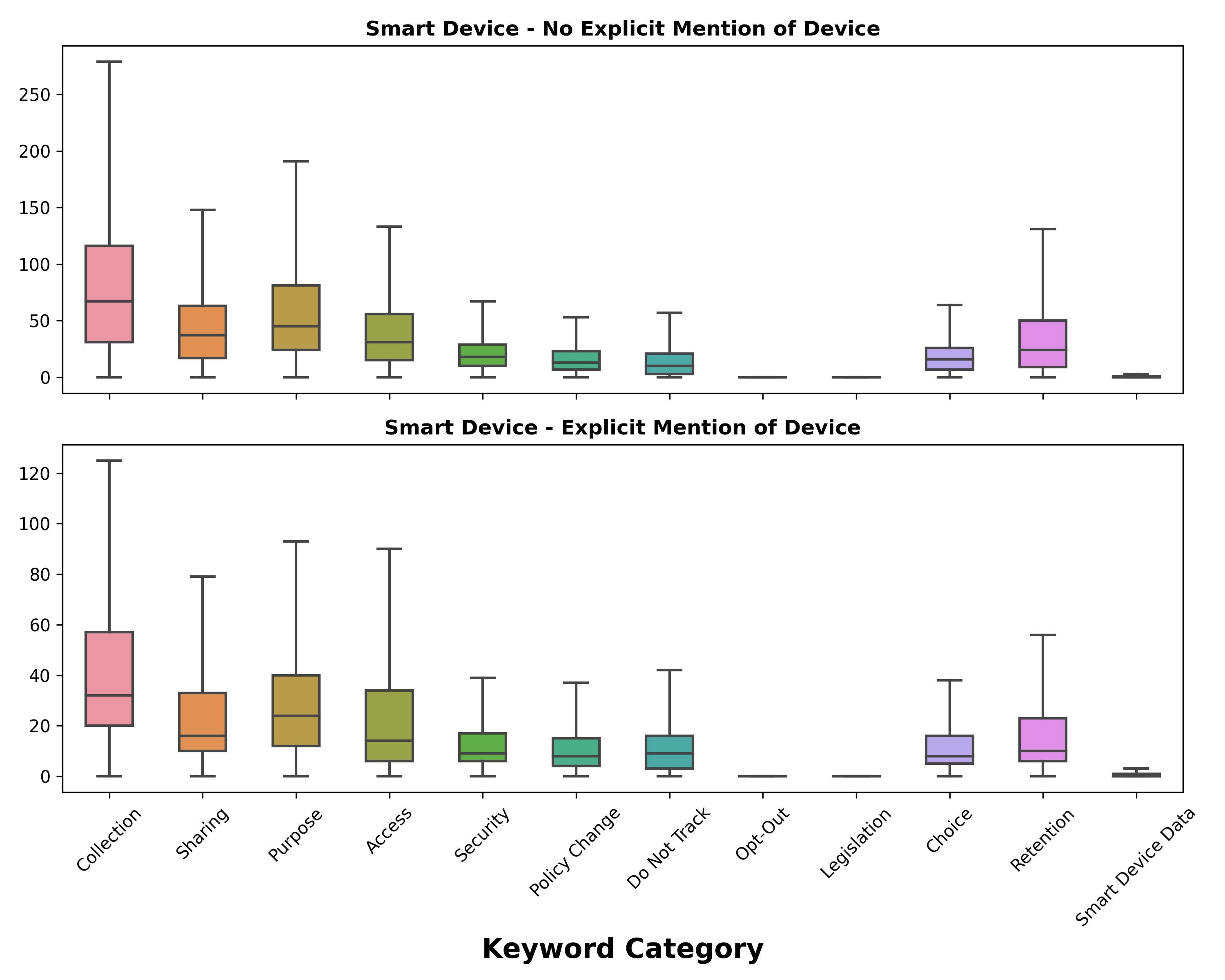}
    \caption{Keyword distribution for smart device policies - With and Without explicit mention of device}
    \label{fig:smartdevices}
\end{figure}

We used sklearn's CountVectorizer to analyze keyword frequencies in smart device policies, comparing those that explicitly mention the devices to those that do not. Our findings, illustrated in Figure \ref{fig:smartdevices}, show that policies without device mentions tend to focus more on general data collection than specifics related to device use. Both policy types consistently address data sharing and purpose. 'Access' and 'Security' keywords are moderately mentioned, indicating a universal emphasis on these aspects. Policies that explicitly mention devices use language related to data access slightly more than those that do not. This greater emphasis on access suggests that these policies are attempting to make the end user aware that the device(s) mentioned require resources that others may not need. However, 'Policy Change' and 'Choice' keywords are less common, suggesting infrequent updates or limited user options in data handling. Less frequent are the 'Do Not Track,' 'Opt-Out,' and 'Retention' keywords, highlighting a potential area for increased policy transparency. 'Legislation' keywords vary slightly, possibly reflecting different legal requirements. Policies mentioning devices specifically tend to more clearly address smart device data. This suggests that smart device privacy policies vary in their focus, with policies not mentioning devices following general data practices while policies not mentioning devices emphasize specific device usage.

\begin{figure}[!htb]
    \centering
    \includegraphics[width=0.8\columnwidth]{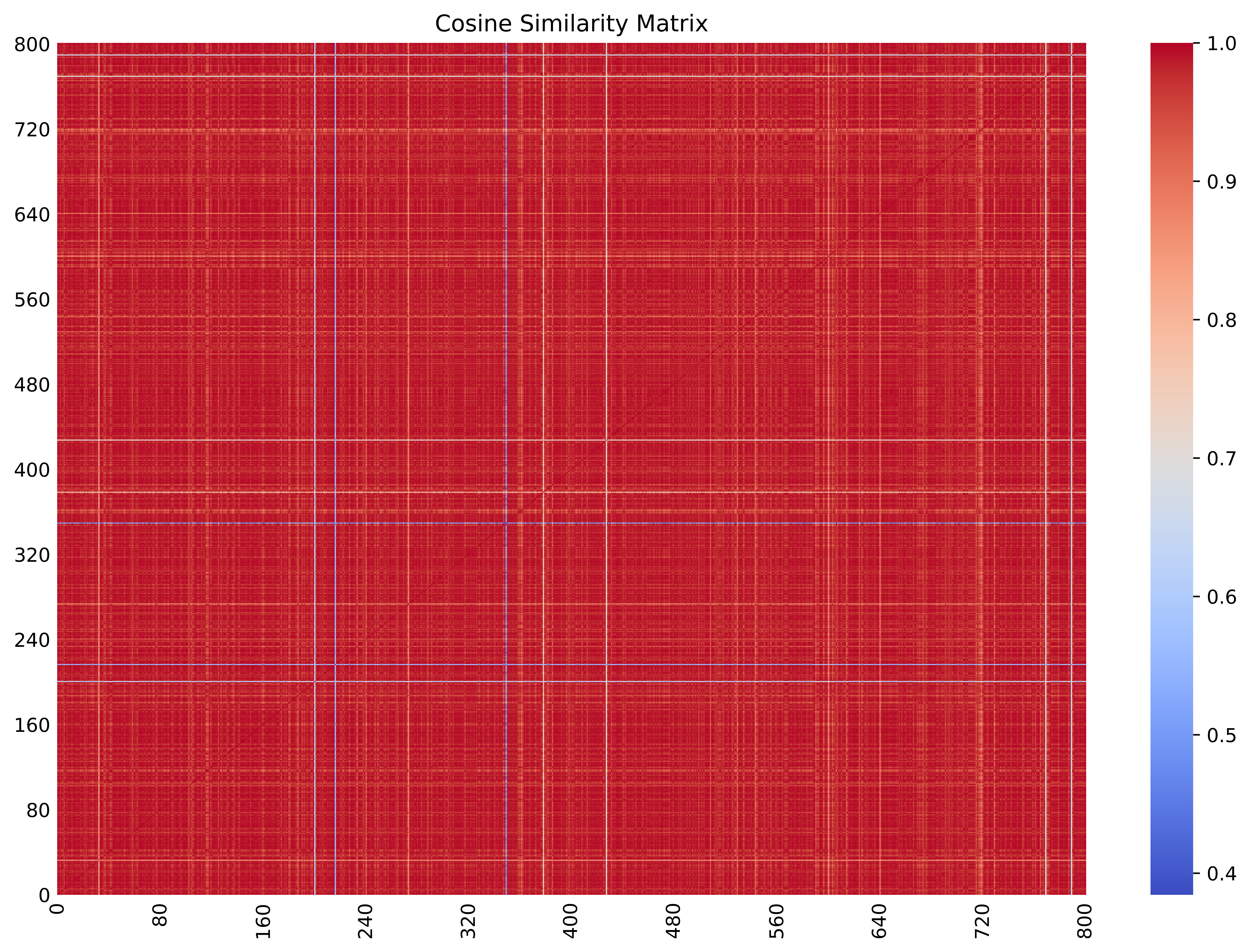}
    \caption{Heatmap of cosine similarities of policy embeddings.}
    \label{fig:heatmap}
\end{figure}

After observing the distribution of keywords between the two subsets of smart device policies, we found that keywords alone cannot fully describe the similarity of the policies. While they provide a baseline understanding, they do not capture the context. To address this, we employed text embeddings to transform the policy documents into numerical vectors, allowing for a more comprehensive and quantifiable comparison of the textual similarity. Cosine similarity, a measure that gauges the cosine of the angle between two vectors, served as our metric for this analysis. Interestingly, there was a striking similarity (94\%) amongst most policies (see \Cref{fig:heatmap}). This suggests the existence of a standard template or common elements that many smart device manufacturers follow while drafting their policies. The remaining 6\% of policies showed significant differences, indicating potential areas of innovation or divergence from common policy templates. However, the presence of some variance, as indicated by lighter shades, reflects the presence of distinctive elements within individual policies. This divergence could be attributed to unique operational practices, targeted user demographics, or specific legal requirements necessitating a departure from the norm. Overall, the high average cosine similarity score implies a cohesive body of privacy policies within the smart device space, albeit with room for individualized approaches.

\subsection{How are the Policies Changing?}

We analyze the evolution of smart device privacy policies based on their changes and their response to the GDPR. 

\begin{figure}[!htb]
    \centering
    \includegraphics[width=\linewidth]{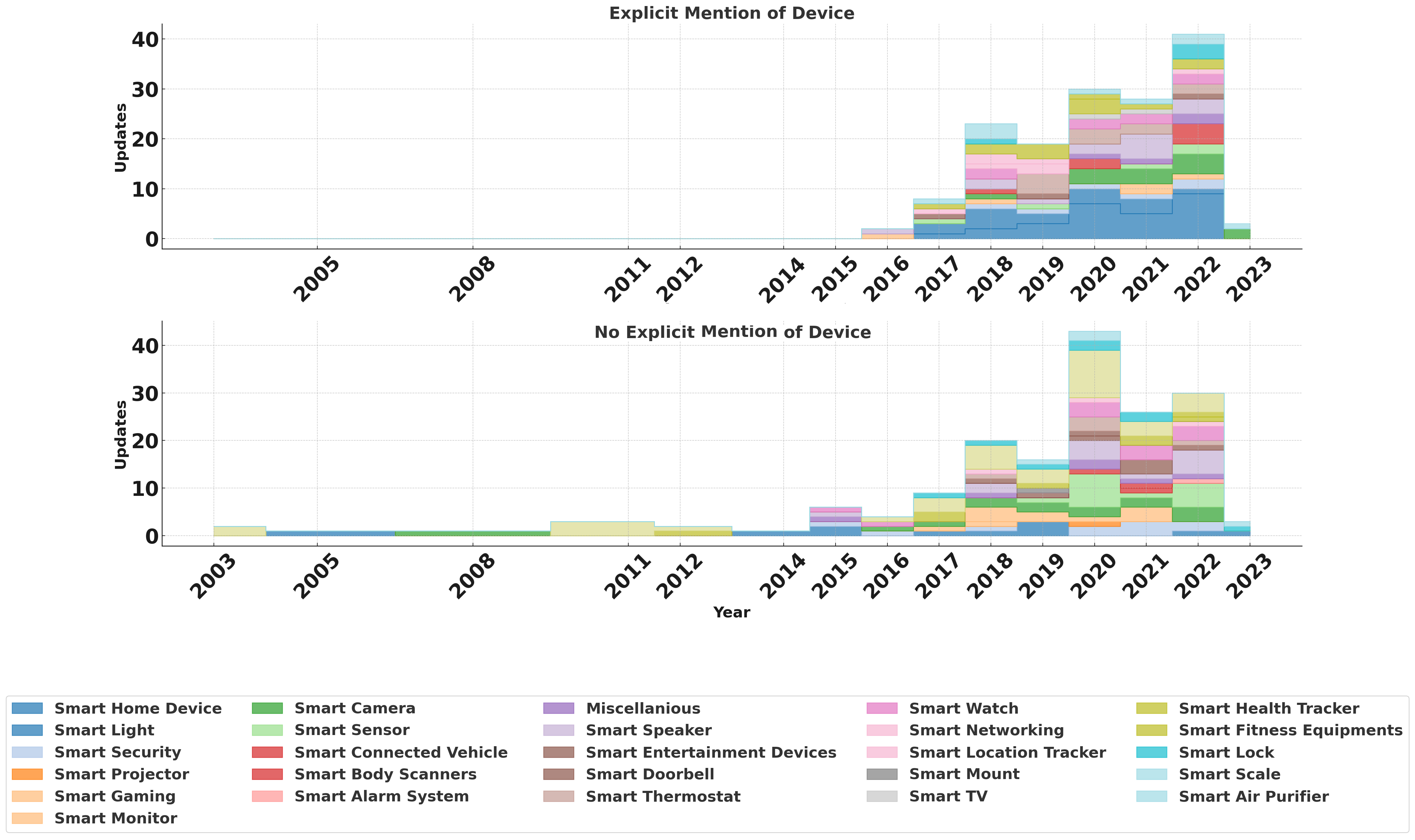}
    \caption{Distribution of privacy policies based on their last update.}
    \label{fig:lastupdate}
\end{figure}

We first analyzed how updated the smart device privacy policies are. \Cref{fig:lastupdate} shows the distribution of privacy policies based on the category of the smart device and their last update. We first observe that only 39.5\% of the policies (324) disclosed their last update date, a concern considering how frequently smart devices receive hardware/software updates. When we separate policies based on whether there is an explicit mention of the smart device, we observe that policies with explicit mentions tend to be more updated for the same category. For instance, for the smart home device category, we note that six policies without explicit mention to the device had a last update before 2013. For nearly 93\% of the policies with an explicit update, the update occurred after the GDPR~\cite{gdpr} became effective in 2018 (63\% were updated after CCPA~\cite{ccpa} became effective in 2020). This suggests that manufacturers acknowledge and adopt the new requirements introduced by these regulations.

\begin{figure}[!htb]
    \centering
    \includegraphics[width=\linewidth]{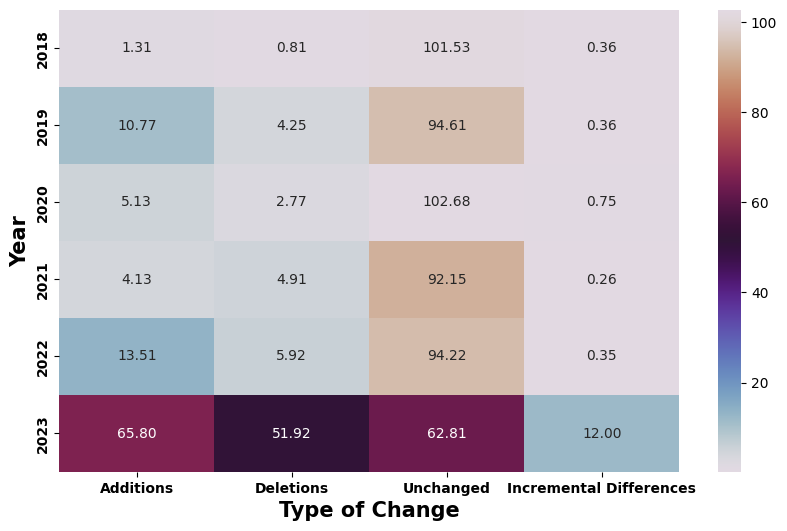}
    \caption{Privacy policy updates by year.}
    \label{fig:heatmapadd}
\end{figure}

Next, we analyze the changes that policies are experiencing in the last 5 years. The heatmap in \Cref{fig:heatmapadd} shows the trends in content modification across different categories: Additions, Deletions, Unchanged, and Incremental Differences. There is a notable uptick in changes between 2018-2019 and 2022-2023. In the more recent timeframe, there is a pronounced spike in Additions and Deletions, averaging 65.80 and 51.92 sentences, respectively, pointing to a significant phase of content restructuring or updates.
Conversely, the Unchanged category displays minimal fluctuation, with a relatively lower average of 62.81 sentences in 2023, suggesting that a portion of the content has consistently remained the same throughout the period. Incremental differences hit a high in 2023 with 12 instances, indicating a tendency towards more detailed and subtle editorial adjustments as time progresses.

\subsection{How Did GDPR Impact Privacy Policies?}

This analysis centers on evaluating the privacy policies of various manufacturers, focusing specifically on changes before and after the implementation of the GDPR in 2018. Out of the 819 extracted policies, we could only obtained archived versions for 312 of them. In the following, we compare the latest archived version before the enactment of GDPR with the oldest archived version after it.

\begin{figure}[!htb]
  \centering
  \includegraphics[width=0.7\columnwidth]{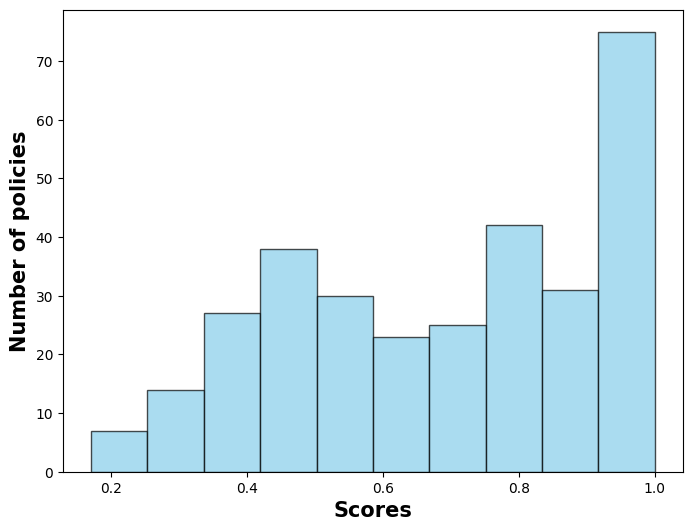}
  \caption{Distribution of similarity scores.}
  \label{fig:image1}
\end{figure}

\begin{figure}[!htb]
  \centering
  \includegraphics[width=\columnwidth]{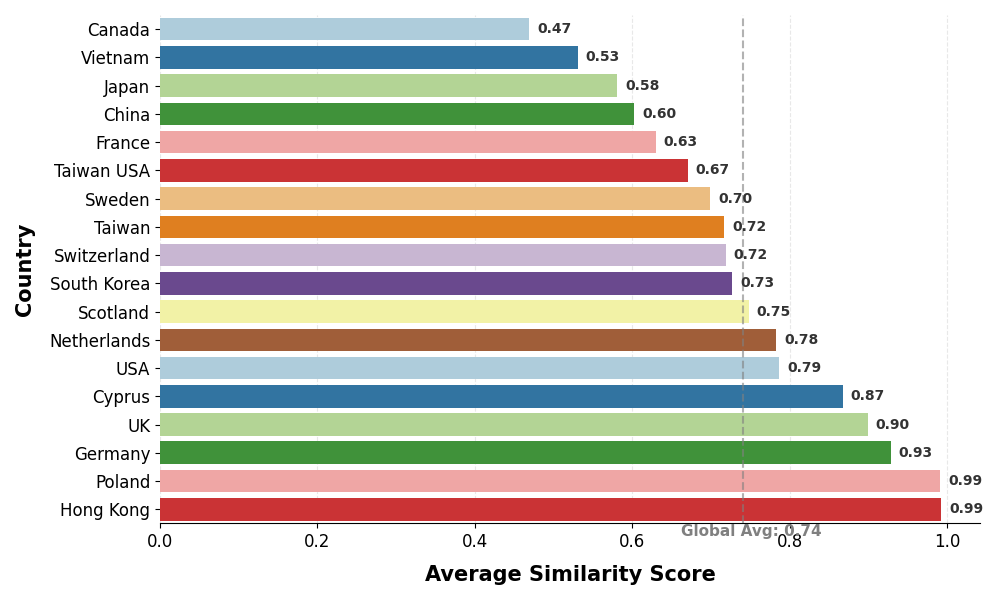}
  \caption{Privacy policy similarity scores across countries.}
  \label{fig:image2}
\end{figure}

\Cref{fig:image1} shows the changes that policies experienced pre/post-GDPR. Note that 75 policies (24\% of the dataset) did not experience any changes and 73 additional policies experienced small modifications (considering a similarity score above 75\%). On the other hand, 14 policies (4.5\%) underwent substantial modifications (their similarity score was below 30\%). This indicates that the majority of the policies were either already in compliance or required only minor adjustments to meet the new standards. Extending the analysis across different countries (see \Cref{fig:image2}) we observe that manufacturers from countries like Canada, Vietnam, Japan, and China significantly changed their policies post-GDPR. In contrast, manufacturers from countries like Germany, the United Kingdom, Poland, and Hong Kong adapted their policies to GDPR well in advance and hence did not require major changes at the time the regulation was enacted. The global average similarity score stands at 0.74 which indicates that most manufacturers did not require significant changes post GDPR.

\begin{figure}[!htb]
  \centering

  \begin{subfigure}[b]{0.48\linewidth}
    \includegraphics[width=\linewidth]{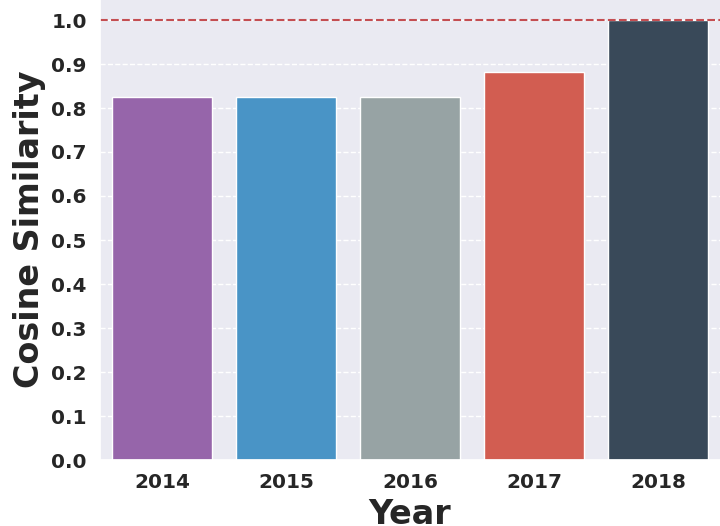}
    \caption{Miele}
    \label{fig:heatmap1}
  \end{subfigure}
  \hfill
  \begin{subfigure}[b]{0.48\linewidth}
    \includegraphics[width=\linewidth]{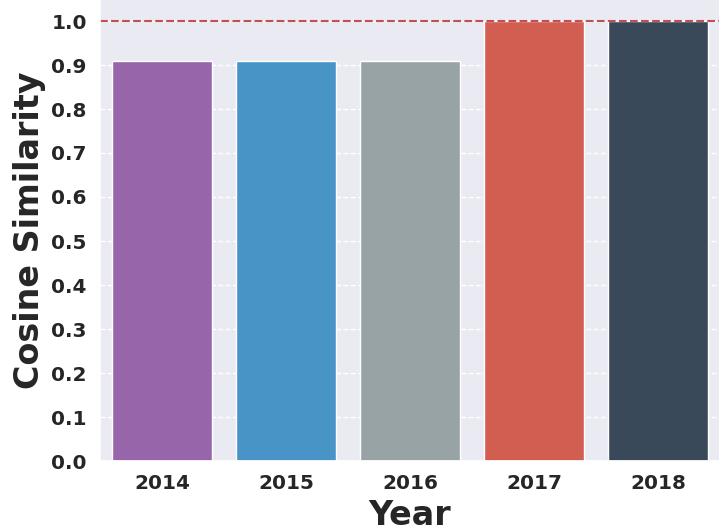}
    \caption{Schluter}
    \label{fig:pca}
  \end{subfigure}
  
  \caption{Similarity across policies for two manufacturers 4 years before GDPR.}
  \label{fig:embeddings}
\end{figure}

As an example of further analysis supported by our dataset, we showcase how the privacy policies of two European manufacturers of smart devices (Miele and Schluter) changed in a 4 years period before the enactment of GDPR. \Cref{fig:embeddings} shows the similarity of a specific year's policy wrt the previous year's policy. In both cases, we observe that policies presented changes during the first three years and, the case of Miele, during the forth year too. Then, they became unchanged after the enactment of GDPR in 2018. This shows that different manufacturers adapted their policies in advance of the regulation.

In conclusion, GDPR has been a significant catalyst for change in smart device privacy policies. The varying degrees and timelines of policy adaptations underscore the regulation's broad impact and the diverse strategies employed by organizations worldwide.


\section{Smart Device Policies vs. Other Domains}
\label{sect:analysis2}




In the following, we explore the similarity of smart device privacy policies vs. those of more consolidated domains of e-commerce and mobile applications. First, we analyze their purely textual similarity using cosine similarity.

\begin{figure*}[!htb]
  \centering
  \begin{subfigure}[t]{.37\textwidth}
    \centering
    \includegraphics[width=\linewidth]{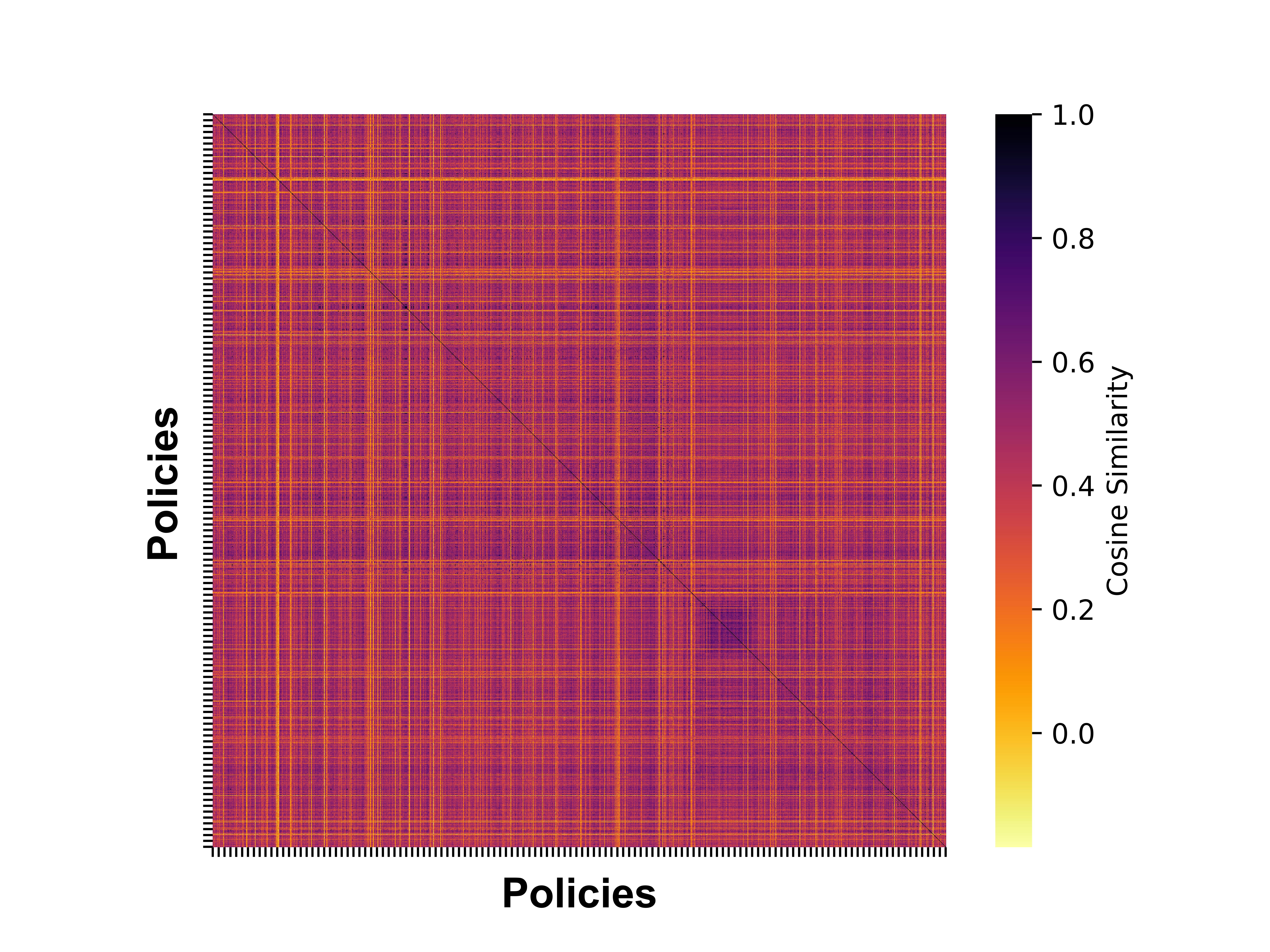}
    \caption{Heatmap}
    \label{fig:heatmap1}
  \end{subfigure}%
  \hfill 
  \begin{subfigure}[t]{.34\textwidth}
    \centering
    \includegraphics[width=\linewidth]{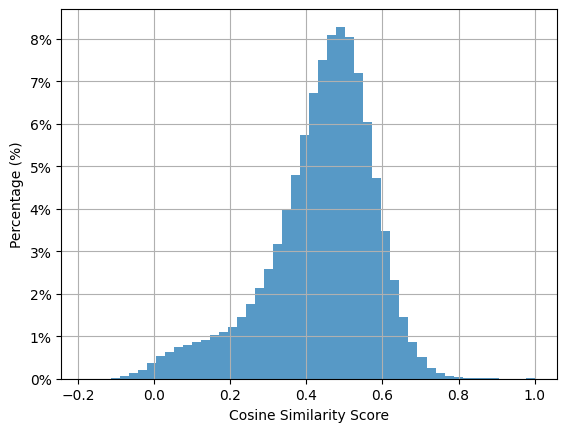}
    \caption{Similarity score distribution}
    \label{fig:distribution2}
  \end{subfigure}
  \caption{Textual similarity across all privacy policies (i.e., smart devices, mobile apps, and e-commerce).}
\end{figure*}

\begin{figure*}[ht]
  \centering
  \begin{subfigure}{.40\textwidth}
    \centering
    \includegraphics[width=\linewidth]{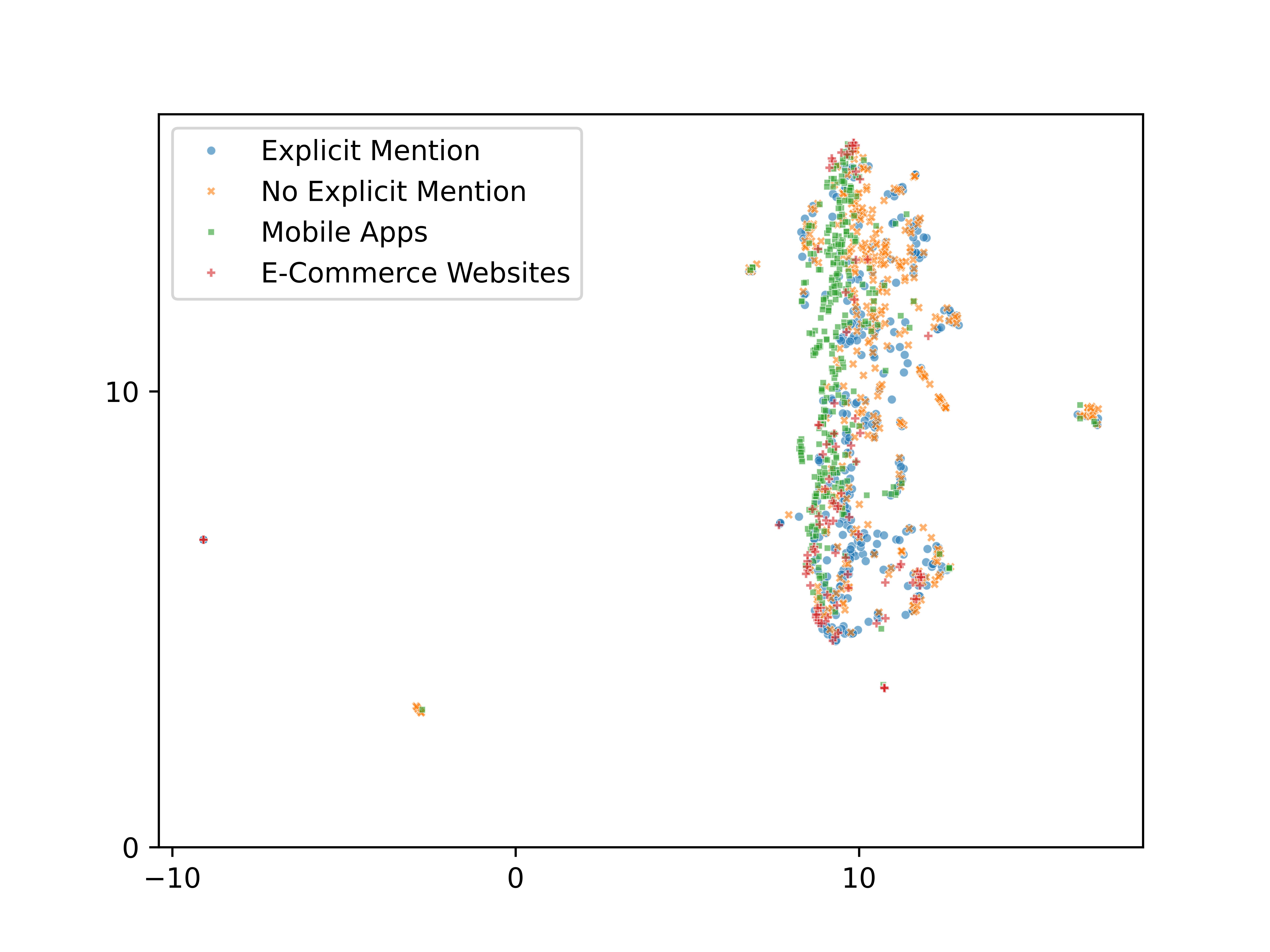}
    \caption{Based on text.}
    \label{fig:scatterplot}
  \end{subfigure}%
  \hfill 
  \begin{subfigure}{.40\textwidth}
    \centering
    \includegraphics[width=\linewidth]{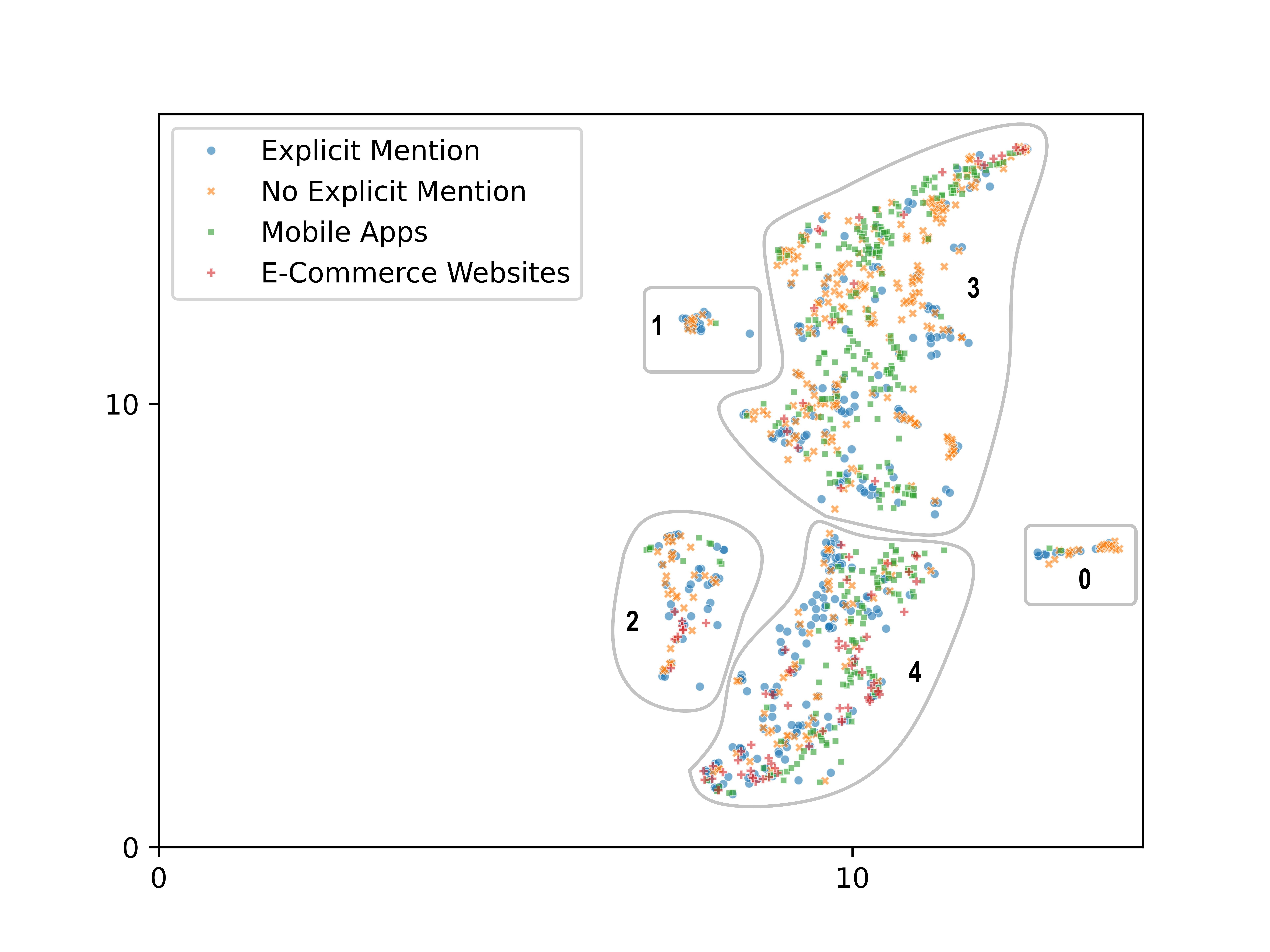}
    \caption{Based on text and features.}
    \label{fig:second_figure}
  \end{subfigure}
  \caption{Policy cluster distributions.}
  \label{fig:clustering}
\end{figure*}

Figure\ref{fig:heatmap1} displays a pairwise cosine similarity analysis of privacy policies, with darker shades indicating higher similarity. The diagonal, being the darkest, reflects perfect similarity as it compares each policy with itself. The off-diagonal elements, showing lighter shades, suggest that while policies share some similarities, there's no complete uniformity. This variation highlights the customization of policies to suit specific data practices of each service. This is also consistent with the similarity analysis performed in Section~\ref{sect:analysis1}.
Figure~\ref{fig:distribution2} presents a histogram of cosine similarity scores, showing a moderate peak with most values between 0.2 and 0.6. This indicates a moderate level of textual similarity across policies. The range of scores suggests some standardization, likely due to legal and regulatory language, but also retains unique elements, possibly due to specific business practices and data-handling requirements of each service.
Finally, we applied HDBSCAN, an efficient clustering algorithm, to group policies based on their text. Figure~\ref{fig:scatterplot} reveals that most policies cluster together, with a few outliers. Notably, policies from smart devices with explicit mentions tend to cluster more closely, indicating more standardized language within this subcategory.

After textual-based analysis revealed a moderate level of textual similarity in privacy policies across smart devices, mobile apps, and e-commerce websites, we used a readability-based analysis to understand the underlying factors contributing to this similarity. Specifically, we sought to understand if readability could account for the medium level of similarity previously identified. Table~\ref{tab:combined} shows the readability analysis results for both mobile apps and e-commerce website policies. When comparing the average mobile app and e-commerce policy with the average smart device policy (see Table~\ref{table:fetaure_avg}), we observe as the main difference reading time: smart device privacy policies are, in general, longer that those of mobile applications but shorter than those of e-commerce websites.

\begin{table}[!htb]
\centering
\resizebox{\linewidth}{!}{
\begin{tabular}{|l|l|l|l|}
\hline
Policy Features & Min Value & Median Value & Max Value \\
\hline
Coherence Score & 0.10 / 0.18 & 0.31 / 0.30 & 0.76 / 0.92 \\
\hline
Freq. of Imprecise Words & 0.00 / 0.00 & 0.02 / 0.02 & 0.04 / 0.03 \\
\hline
Freq. of Connective Words & 0.00 / 0.00 & 0.04 / 0.04 & 0.07 / 0.07 \\
\hline
Reading Complexity & 6.44 / 7.02 & 12.92 / 12.37 & 23.09 / 26.68 \\
\hline
Reading Time (Min) & 1.00 / 1.00 & 8.00 / 22.00 & 185.00 / 205.00 \\
\hline
Entropy & 6.24 / 6.06 & 7.87 / 8.27 & 9.07 / 9.71 \\
\hline
Freq. of Unique Words & 0.07 / 0.07 & 0.28 / 0.19 & 0.60 / 0.54 \\
\hline
Grammatical Errors & 0.00 / 0.00 & 0.00 / 0.007 & 0.12 / 0.01 \\
\hline
\end{tabular}
}
\caption{Readability analysis statistics for privacy policies of mobile apps (first value before the '/') and e-commerce websites (second value).}
\label{tab:combined}
\end{table}

\begin{figure}[!htb]
  \centering
  \includegraphics[width=\columnwidth]{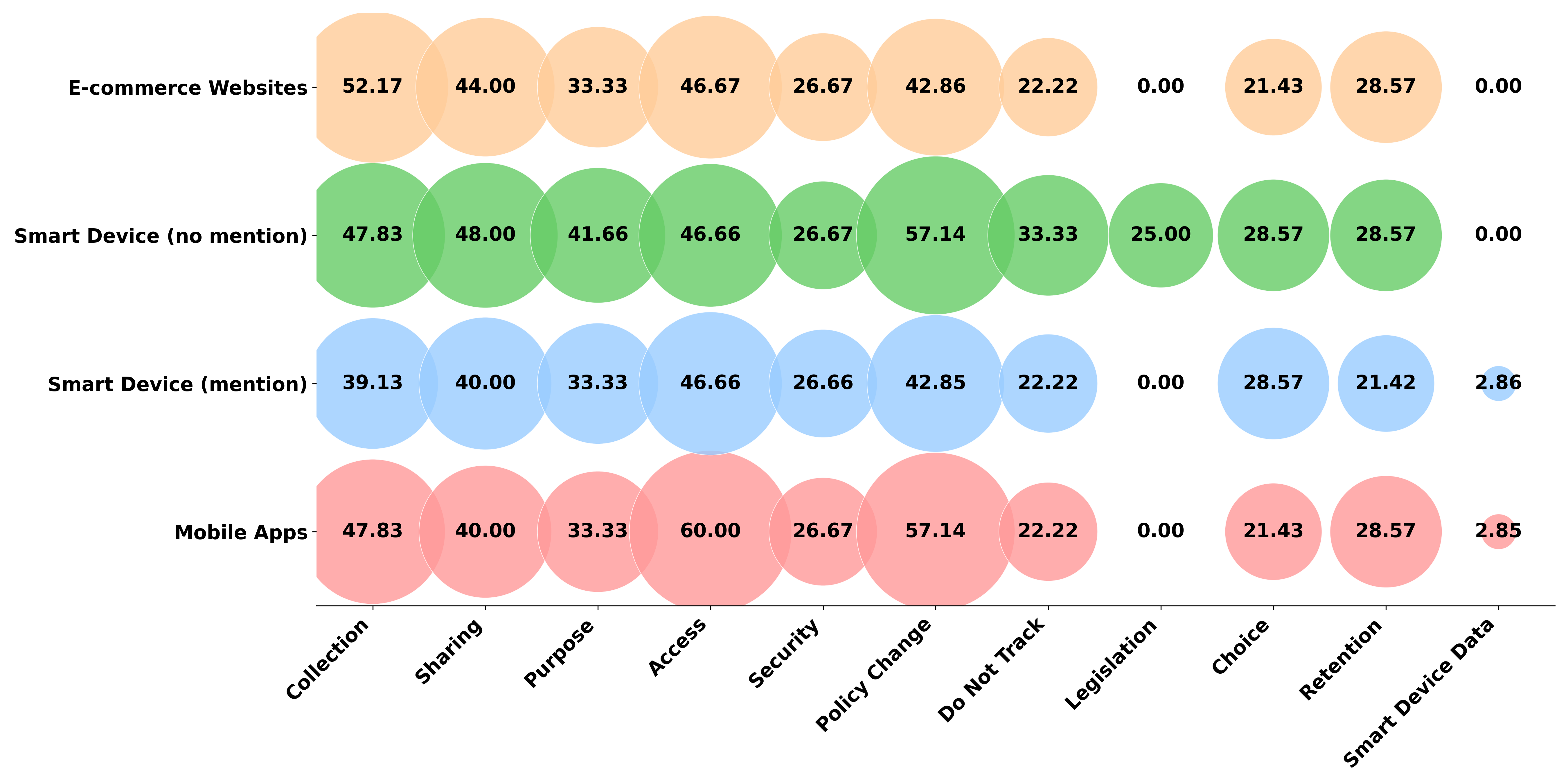}
  \caption{Distribution of keyword coverage.}
  \label{fig:privacy_keywords}
\end{figure}

Next, we analyzed the privacy policies based on extracted privacy features related to data management practices. Figure \ref{fig:privacy_keywords} shows the keyword distribution across categories for e-commerce and mobile apps. Compared to the distribution for smart devices in the previous section (see Figure~\ref{fig:smartdevices}), we observe that mobile apps and smart devices with explicit device mentions have similar privacy policy distributions. This similarity could stem from the fact that mobile apps often use device sensors, akin to those in smart devices. Conversely, e-commerce websites' privacy policies more closely resemble those of smart devices without explicit device mentions. This parallel might be because these policies typically focus on the data management practices of the manufacturer's website rather than the device itself.

\begin{figure}[!htb]
    \centering
    \includegraphics[width=\linewidth]{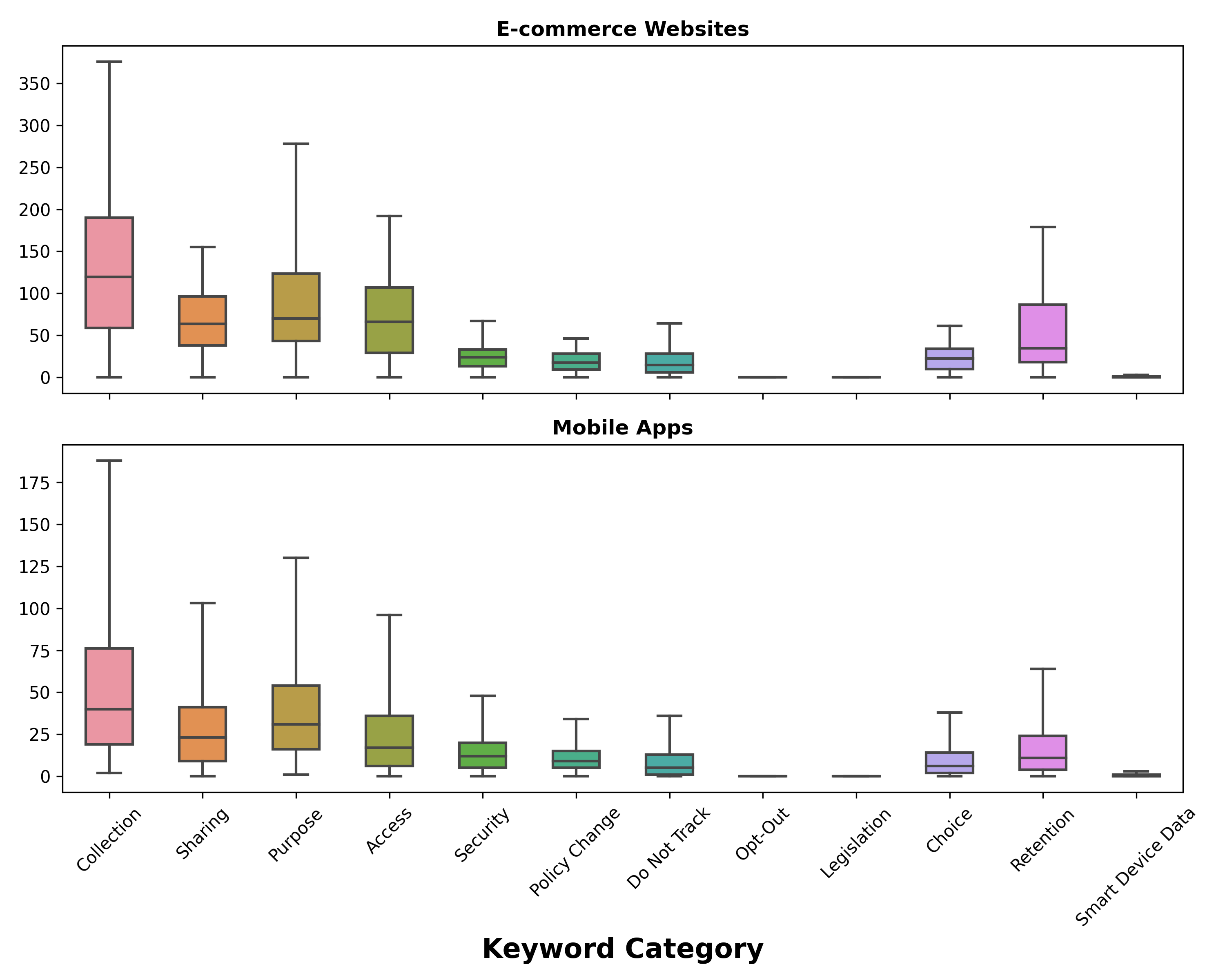}
    \caption{Keyword distribution for e-commerce and mobile apps.}
    \label{fig:ecommerce_mobileapps}
\end{figure}

The comparative bubble chart (see Figure \ref{fig:privacy_keywords})  clearly illustrates the trends in privacy policy attributes (see Table \ref{tab:privacy_keywords}) for e-commerce websites, smart devices, and mobile apps, providing a percentage-based analysis of keyword coverage. It reveals a nearly equal focus on data collection policies between websites (52.17\%) and Android apps (47.83\%), while IoT policies less frequently mention devices. Data sharing is a significant concern in all areas, particularly in IoT policies without device mentions (56\%), indicating a trend towards more data-sharing practices. Notably, 'Do Not Track' and 'Choice' are prominent in the general IoT category, suggesting a broader privacy approach. In contrast, specific IoT device policies seem to have a more focused privacy strategy. This, along with the varied representation of 'Access' and 'Choice', suggests a fragmented approach to user empowerment in managing their data. Android policies often clarify access purposes (60.\%), reflecting a commitment to transparency, while IoT device policies commonly detail policy changes (57.14\%), indicating a move towards enhanced user control. 'Choice' is widely recognized, especially in the IoT (Device Not Mentioned) category (35.71\%). Policy updates are frequently communicated in Android environments (57.14\%), reflecting a proactive approach. However, 'Legislation' is less often addressed, with the highest mention in the general IoT category (25\%), pointing to variations in legal compliance. The prevalence of 'Purpose' in almost all policies (33.33\%) suggests it's becoming an industry norm. Data retention is uniformly emphasized across domains (28.57\%), but IoT-specific data concerns are minimally discussed, with the most in policies for specific IoT devices (2.86\%), indicating a potential area for improvement in policy development.

Figure \ref{fig:ecommerce_mobileapps} complements this by illustrating the spread and central tendency of keyword mentions across different categories, providing insight on the emphasis placed on various aspects of privacy. For instance, keywords such as 'Collection' and 'Purpose' display a higher median in Android policies, suggesting a more frequent mention that may indicate a heightened focus on these areas within the platform. In contrast, 'Do Not Track' and 'Legislation' appear less frequently addressed across all datasets, as denoted by lower medians and many outliers, pointing to their sporadic mention and raising questions about the commitment to tracking transparency. In the case of IoT devices mentioning specific devices, there is a lower median frequency for most keywords, implying a more targeted or concise approach to their privacy policies. IoT devices that do not mention specific devices have a higher median frequency. Despite variations, there is a standardization level in the policies of three datasets, particularly those that mention the device, as evidenced by the tighter interquartile ranges. Finally, by integrating all the features from our textual and readability analyses into a multi-dimensional clustering approach, we observed intricate patterns of similarities and dissimilarities among the privacy policies.


The scatter plot (see Figure \ref{fig:second_figure}) elucidates the grouping of policies based on readability and privacy-related features, along with policy text revealing clusters that vary in coherence, complexity, linguistic precision, etc. Cluster 0 presents a range of reading complexities, merging technical and user-friendly documents, highlighting variability in user understanding. These policies mainly focus on data collection, as evidenced by frequent "Collection" keyword usage. Cluster 1 is characterized by coherent, simpler policies, indicative of a user-friendly approach, and emphasizes "Collection", "Purpose", and "Retention" keywords. Cluster 2, with the widest coherence and simplest readability, suggests clear, straightforward policies, likely enhancing user comprehension. This cluster frequently uses "Collection", "Sharing", "Purpose", and "Access" keywords. Cluster 3 comprises policies with long reading durations and higher complexity, pointing to comprehensive, yet possibly overwhelming content, showing low to moderate keyword usage across categories. Finally, Cluster 4 strikes a balance in detail and readability, though with variable reading times and grammatical precision, potentially affecting user experience, while exhibiting the highest average keyword usage across all categories.

Overall, the clustering analysis confirms that privacy policies exhibit moderate textual similarities but display diverse readability and privacy-related characteristics. This diversity reflects the balance between the necessity for standardized legal language to ensure compliance and specificity tailored to the unique requirements of different service types. The varied clustering underscores the challenges users face in navigating privacy policies, with some documents being more user-centric and others demanding higher levels of literacy and technical understanding.

\section{Discussion}
\label{sect:discussion}
We would like to highlight several important observations from our study. 

Our analysis reveals the challenges that today's customers might face in accessing information about the data management practices of smart device products. This would make it difficult, and in some cases even impossible, to make an informed decision based on privacy at the time of purchasing certain smart devices.

\begin{itemize}
    \item Notably, popular e-commerce platforms like Amazon and Walmart often do not provide direct links to the privacy policies of listed smart devices and/or information about their data management practices. Consequently, customers may need to search for their policies, typically on the manufacturers' websites. However, our analysis found that a significant portion of popular smart device manufacturers (1,167, 53\% of the total) do not have websites beyond their Amazon store pages. This is particularly concerning since some of those manufacturers sell smart devices such as baby monitors and microphone-enabled speakers which handle sensitive information. To assess these manufacturers, we used the popular website \textit{Fakespot}~\cite{fakespot}, which uses AI to detect fake reviews and potential scams. Of the 572 manufacturers analyzed by Fakespot, 52\% received a 'C' grade or lower, suggesting that nearly half (the ones obtaining a higher grade) have real customers leaving honest reviews (see Figure~\ref{fig:fakespot_review} for more details). This situation underscores the need for e-commerce sites to require smart device manufacturers to include detailed information about their privacy policies alongside their products, enhancing transparency for consumers.

    \item Moreover, even when a customer locates a smart device manufacturer's privacy policy, many of these documents lack clarity regarding their applicability. Our analysis showed that only 49\% of the analyzed policies explicitly state they pertain to the smart device in question. This lack of specificity can lead to misconceptions, such as assuming a smart speaker does not record voice data or a smart camera does not capture video, simply because audio or video collection is not mentioned in the policy. Therefore, it is crucial for smart device manufacturers to enhance the transparency of their privacy policies. 

    \item Although the average smart device privacy policy is relatively readable, there are instances where the complexity and required reading time render it impractical for most customers to fully grasp the content. This issue is not unique to smart devices; we observed similar challenges with mobile apps and e-commerce websites. However, the concern is more pronounced for smart devices due to the potentially sensitive data they collect via their equipped sensors, emphasizing the need for clearer and more accessible privacy policies in this sector.
\end{itemize}

Our study highlights the positive influence of data protection regulations like GDPR on the clarity of smart device privacy policies, a trend also seen in other domains. Post-GDPR, we noticed changes in these policies, leading to reduced ambiguity. Additionally, our comparison with more established domains reveals a notable uniformity in the structure and language of all privacy policies, including those of smart devices. This uniformity, while indicative of a 'template' approach, could be a double-edged sword. On one hand, it promotes standardization; on the other, it might lead to a lack of specific details tailored to different user needs and contexts. This generalized approach could result in users overlooking critical information about the sensitive data collected by smart devices.


\section{Conclusions and Future Work}
\label{sect:conclusion}
Smart devices, increasingly prevalent in daily life, pose significant privacy challenges due to their potentially intrusive sensors, which permeate personal spaces like homes. Our research presents, to our knowledge, the first comprehensive analysis of smart device privacy policies. Building on existing methodologies primarily used to analyze privacy policies of e-commerce websites and mobile apps, we developed a framework to discover, collect, and analyze the privacy policies of smart devices.
Our framework addresses the challenge of locating these policies by searching e-commerce platforms for smart devices and identifying their manufacturers. It then seeks out the relevant privacy policies on the manufacturers' websites. The framework incorporates a suite of advanced machine learning and natural language processing techniques to derive insights from these policies. This includes evaluating readability, identifying privacy-related topics, and tracking their historical evolution. We have made both the tool and its source code publicly available, fostering further research in this area.

Our analysis encompassed over 4,000 smart devices, representing a significant portion of those sold on e-commerce websites, across 28 categories. It involved over 2,000 smart device manufacturers and examined 1,000 privacy policies. Additionally, we compared these privacy policies with those in the established domains of e-commerce and mobile applications. 
As a future work, we aim to enhance our tool to include user reviews that specifically mention privacy concerns related to smart devices. We also plan to integrate data on cybersecurity incidents and data breaches. This expansion will enable more comprehensive studies on the topic, providing deeper insights into the privacy and security landscape of smart devices.



\bibliographystyle{unsrt}
\bibliography{bibliography.bib}


\section*{}
\label{sect:appendix}
\clearpage
\begin{appendices}

\section{Privacy Policy Analysis Details}

\Cref{table2} contains the taxonomy of imprecise words and \Cref{conn} contains the taxonomy of connective words (both extracted from~\cite{kotal2021effect}) that \system uses in its analysis of an IoT device privacy policy.

\begin{table}[!ht]
\small
\centering
\resizebox{\linewidth}{!}{
 \label{tab:title} 
\begin{tabular}{|c|c|p{1cm}p{1cm}p{1cm}|}
\hline
\multicolumn{4}{|c|}{\textbf{Imprecise Words}}\\
\hline
&\multicolumn{3}{|c|}{may,might,likely,can}\\\textbf{Modal Words}&\multicolumn{3}{|c|}{could,would}  \\ \hline
&\multicolumn{3}{|c|} {easy,adaptable}\\\textbf{Usable Words}&\multicolumn{3}{|c|}{familiar,extensible} \\ \hline
\textbf{Probable Words}&\multicolumn{3}{|c|} {probably,possibly,optionally} \\ \hline
&\multicolumn{3}{|c|} {anyone, certain}\\&\multicolumn{3}{|c|}{everyone, numerous}\\
\textbf{Numeric Words}&\multicolumn{3}{|c|}{some,most,few}\\&\multicolumn{3}{|c|}{much,many,various}\\
      &\multicolumn{3}{|c|}{including but not limited to
} \\ \hline
&\multicolumn{3}{|c|} {depending,necessary}\\
&\multicolumn{3}{|c|}{inappropriate,appropriate}\\\textbf{Condition Words}&\multicolumn{3}{|c|}{as needed,as applicable}\\
      &\multicolumn{3}{|c|}{ otherwise reasonably}\\&\multicolumn{3}{|c|}{from time to time}
      \\ \hline
&\multicolumn{3}{|c|} {generally,mostly,widely}\\
&\multicolumn{3}{|c|}{commonly,usually,general}\\\textbf{Generalization }&\multicolumn{3}{|c|}{Normally,typically,largely}\\
\textbf{Words}&\multicolumn{3}{|c|}{often,primarily}\\
      &\multicolumn{3}{|c|}{among other things}
      \\ \hline
\end{tabular}
}
\caption{Taxonomy of imprecise words extracted from~\cite{kotal2021effect}.}
\label{table2}
\end{table}

\begin{table}[!htb]
\centering
\resizebox{\linewidth}{!}{
\begin{tabular}{|l|l|}
\hline
\multicolumn{2}{|c|}{\textbf{Connective Words}}\\
\hline
\bf Copulative Words&and, both, as well as,
not only, but also\\
\hline
\bf Control Flow Words& if, then, while  \\
\hline
\bf Anaphorical Words& it, this, those  \\
\hline
\end{tabular}
}
\caption{Taxonomy of connective words extracted  from~\cite{kotal2021effect}.}
\label{conn}
\end{table}

\section{Seed Keywords}
\begin{table*}[!htb]
\centering
\footnotesize
\begin{tabularx}{\textwidth}{|p{2.8cm}|X|p{6cm}|}
\hline
\textbf{Privacy Attribute} & \textbf{Definition} & \textbf{Keywords} \\
\hline
Collection & The methods and purposes used by a service provider to get user data. & Collect, personal, identifiable, telephone, number, phone, telephone number, IP address, phone number, IP, mobile, email,
address, name, date of birth, birthday, age, account, credit card, location, username, password, contact, zip code, postal code,
mailing address, phone \\
\hline
Sharing& The methods third parties use to share or acquire user information. & Party, share, sell, disclose, company, advertiser, provider, partner, public, analytics, companies, organizations, businesses,
contractors, divulge, sell, law, legal, regulation, third party, transfer, service providers, marketing partners, subsidiaries,
disclosure, safe harbor \\
\hline
Purpose& The objectives and reasons behind collecting and using user data.&Ads, use, services, verifying, purpose, fraud, prevention, improve products, identification, promotions, personalize, advertising,
analytics\\
\hline
Access & If users may access, edit, or remove their information, and how. &Delete, profile, correct, account, change, update, section, access, removal, request, modify, edit, settings, preferences, accurate \\
\hline
Security & How user data is safeguarded. & Secure, security, safeguard, protect, compromise, encrypt, advertiser set, unauthorized, access, SSL, socket, socket layer,
encryption, restrict, fraud \\
\hline
Policy Change & Whether and how users will be informed of privacy policy changes. &  Change, change privacy, policy time, current, policy agreement, update privacy, update, notice \\
\hline
Do Not Track & Whether and how internet tracking and advertising using Do Not Track signals are handled. &Signal, track, track request, browser, disable, track setting, cookies, web beacons, IP address \\
\hline
Legislation & The legal frameworks that empower individuals to control the collection, usage, and distribution of their personal information by businesses and organizations. & GDPR, CCPA, General Data Protection Regulation \\
\hline
Choice & User's right to make decisions about how their personal data is collected, used, and shared by a service or platform. & Opt, unsubscribe, disable, choose, choice, consent, setting, option, wish, agree, opt-in, opt-out, subscribe, do not track \\
\hline
Retention &The policies and practices related to the storage, archiving, and deletion of user data   &Retain, store, delete, deletion, database, participate, promotion, send friend, record, remove, retention, keep, data, backup,\\
\hline

Smart Device Data & Information collected from IoT devices and related services. & sensor data, device data, environmental data, operational data, health metrics, location data, Bluetooth, Wi-Fi, NFC, Capacitive NFC, LTE, device communication, connected devices, smart devices, user commands, voice control, device settings, interaction logs, health data, fitness tracking, biometric identifiers, heart rate, proximity data, ambient conditions, temperature, lighting, microphone, Barometer, cross-device tracking, RFID, Vibration sensor, Radar sensor, Pressure sensors, Ultrasonic sensors, Infrared sensors, RF sensor \\
\hline
\end{tabularx}
\caption{Privacy attributes extracted from a privacy policy based on seed keywords.}
\label{tab:privacy_keywords}
\end{table*}

Table\ref{tab:privacy_keywords} contains the seed keywords for evaluating data practices in privacy policies. we evaluated privacy policy sections based on established best practices in the field. These sections include Collection, which details the types and methods of user data collection; Sharing, explaining how and with whom user data is shared; Choice, presenting users' privacy options such as opting in or out; Access, outlining how users can access and verify their data; Data Retention, which describes the reasons and duration for storing user data; Data Security, focusing on the measures taken to protect user data; Policy Change, explaining the procedure for notifying users about changes in privacy practices; Do Not Track, which concentrates on online tracking technologies like cookies; and Purpose, clarifying the intended use of collected data and ensuring it is not used for unapproved purposes. Each section is crucial for understanding how user data is managed and protected

\section{Framework Evaluation}
In this section, we assess the effectiveness and performance of \system by conducting an evaluation of both its policy collection and analysis capabilities.

\subsection{Evaluating Policy Collection}

\paragraph{Current Privacy Policy Extraction.}
We evaluated the accuracy of our Amazon web-scraped data against verified "truth" values for the top 30 products across ten smart device categories, totaling 300 devices. Our assessment primarily focused on extracting manufacturers' names and their websites, using three key metrics: F1 score, recall, and precision. The results, detailed in \Cref{tab:collection_results}, show a high overall F1 score (0.98), recall (0.96), and precision (0.99) for manufacturer data. Similarly, for website URL collection, the system also achieved impressive scores: an F1 score of 0.95, recall of 0.91, and precision of 0.99. These findings demonstrate the system's effectiveness in accurately extracting essential information about IoT device manufacturers and their websites, a critical step in locating their privacy policies.

\begin{table}[ht]
\centering
\small
\begin{tabular}{lccc|ccc}
\hline
 & \multicolumn{3}{c|}{\textbf{Manufacturer Collection}} & \multicolumn{3}{c}{\textbf{Website Collection}} \\
\textbf{Category} & \textbf{Recall} & \textbf{Prec.} & \textbf{F1} & \textbf{Recall} & \textbf{Prec.} & \textbf{F1}\\
\hline
Sensor & 0.90 & 1.00 & 0.95 & 0.95 & 1.00 & 0.98 \\
Projector & 1.00 & 1.00 & 1.00 & 0.94 & 0.94 & 0.94 \\
Bulb & 1.00 & 1.00 & 1.00 & 1.00 & 1.00 & 1.00 \\
Speaker & 0.97 & 1.00 & 0.98 & 0.97 & 1.00 & 0.98 \\
Alarm & 0.97 & 1.00 & 0.98 & 0.89 & 1.00 & 0.94 \\
Camera & 0.97 & 1.00 & 0.98 & 0.85 & 1.00 & 0.92 \\
Scale & 0.90 & 1.00 & 0.95 & 0.92 & 1.00 & 0.96 \\
Watch & 1.00 & 1.00 & 1.00 & 1.00 & 1.00 & 1.00 \\
Lock & 0.97 & 1.00 & 0.98 & 0.88 & 1.00 & 0.93 \\
Tracking & 0.97 & 0.97 & 0.97 & 0.74 & 1.00 & 0.85 \\
\hline
Overall & 0.96 & 0.99 & 0.98 & 0.91 & 0.99 & 0.95 \\
\hline
\end{tabular}
\caption{Evaluation of the extraction of manufacturers (Manufacturer Collection) and their websites (Website Collection) for IoT devices.}
\label{tab:collection_results}
\end{table}

To evaluate the accuracy of our method in distinguishing IoT products from a broader product range, we randomly selected 100 products from our total dataset and manually verified each to determine if it was an IoT device, establishing a baseline of true identifications. We then applied our method to these same products to classify them as IoT or non-IoT, generating a set of predictions. By comparing these predictions with the manual identifications, we assessed the effectiveness of our method, primarily using the F1 score as our metric. Our approach achieved an F1 score of 0.88, indicating a high level of accuracy in correctly identifying IoT products and effectively minimizing misclassification of non-IoT products as IoT. This result demonstrates the robustness of our method in accurately filtering and identifying IoT products.

To assess the efficacy of \system in collecting privacy policies, we randomly chose 100 manufacturer websites from our dataset and manually identified the privacy policy URL on each site. We then compared these URLs with those automatically extracted by \system. The system demonstrated strong performance, achieving an F1 score of 0.94 in identifying privacy policy URLs and a 0.76 F1 score in extracting policy text. This indicates \system's high precision and accuracy in extraction tasks. However, \system faced challenges when privacy policies were embedded in dynamic website components or required downloading, highlighting a need for specialized parsers for certain sites. Overall, these results suggest that \system's web parsing technique is effective for collecting privacy policies.

\paragraph{Past Privacy Policy Extraction.}
In our study, we randomly selected another 100 manufacturers, ensuring they existed as of 2020, and their corresponding websites from our dataset. We then chose a random date between 2020 and 2022 for each and used \system to retrieve the archived website version from that date, if available. \system successfully retrieved archived websites for 64\% of these manufacturers. However, upon manual review, we found that 8\% of these retrieved sites led to empty homepages, likely due to incomplete captures by the Wayback Machine. Further analysis revealed that \system located the archived privacy policy on the websites of 30 out of the 64 manufacturers (47\%). The failure to find privacy policies for the remaining 44\% was primarily due to missing privacy links in the snapshots. Additionally, the average time difference between the requested and actual snapshot dates was 87 days.



\subsection{Evaluating Policy Analysis}
\label{appendix:a}

\begin{table}[!h]
\centering
\begin{tabular}{|l|c|}
\hline
\textbf{Privacy Document} & \textbf{Value (Our Approach)} \\
\hline
Minimum Correct Grammar & 0.00 \\
\hline
Minimum Imprecise Words & 0.00 \\
\hline
Minimum Connective Words & 0.02 \\
\hline
Maximum Correct Grammar & 0.95 \\
\hline
Maximum Imprecise Words & 0.04 \\
\hline
Maximum Connective Words & 0.06 \\
\hline
\end{tabular}
\caption{ Approach}
\label{tab:our_approach_label}
\end{table}

\begin{table}[!h]
\centering
\begin{tabular}{|l|c|}
\hline
\textbf{Privacy Document} & \textbf{Value (Ground Truth)} \\
\hline
Minimum Correct Grammar & 0.06 \\
\hline
Minimum Imprecise Words & 0.09 \\
\hline
Minimum Connective Words & 0.025 \\
\hline
Maximum Correct Grammar & 0.23 \\
\hline
Maximum Imprecise Words & 0.09 \\
\hline
Maximum Connective Words & 0.076 \\
\hline
\end{tabular}
\caption{ Ground Truth}
\label{tab:ground_truth_label}
\end{table}

To evaluate the insights extracted from each privacy policy, we leveraged existing ground truth datasets from the literature. As there does not exist a human-annotated dataset of privacy policies for IoT devices, we chose datasets of privacy policies of websites that include annotations. The dataset in~\cite{kotal2021effect}, which is based on the OPP-115 dataset~\cite{peters2018deep}, contains annotations on grammatical errors, frequency of imprecise words, and connectivity words. Then, for the remaining features of \system's readability analysis, we annotate 10 policies of the OPP-115 using popular and free web-based resources: \textit{Readable}\footnote{\url{https://readable.com}} for readability, \textit{Planetcalc}\footnote{\url{https://planetcalc.com}} for entropy, and \textit{The Read Time}\footnote{\url{https://thereadtime.com/}} for reading time. We show the results next for each of the features.

W.r.t. \textbf{grammatical errors}, we adhered to the methodology of the benchmark study~\cite{kotal2021effect} for our comparative analysis (see Tables \ref{tab:our_approach_label} and \ref{tab:ground_truth_label}). Despite the benchmark's varied results, we effectively compared our specific scores. Our results for the policy with the fewest grammatical errors closely matched their lowest range, with a minor deviation of -0.06. However, in the policy with the most errors, we noted a significant variance, surpassing their range by approximately +0.7. This suggests that our grammatical error detection tool, \textit{language\_tool\_python}, adopts a more rigorous approach to grammatical correctness.

For \textbf{connective words}, we aligned our findings with the general ranges provided in the study~\cite{kotal2021effect}. The policy with the fewest connective words showed a negligible deviation of 0.005, closely matching their lower range. In contrast, the policy with the most connective words had our figures slightly exceeding theirs by 0.016, yet still demonstrating consistency with the reported study.

Regarding \textbf{imprecise words}, our analysis of the usage in policies compared to the ranges from the study~\cite{kotal2021effect} revealed that our results were slightly higher in the policy with the most imprecise words, by a difference of 0.05. Conversely, for the policy with the fewest imprecise words, our findings were marginally lower, with a difference of 0.09. This discrepancy still underscores the alignment of our method with the findings of the cited study.

Our Flesch-Kincaid \textbf{readability} scores closely aligned with those from Readable, indicating methodological agreement. For example, for RedOrbit's privacy policy, our score was 10.26 compared to Readable's 10.4, and for Sci-News, our score was 9.4 against Readable's 8.9. Similar trends were observed for Uptodate and Earthkam, where our scores were 12.9 and 14.6, respectively, in comparison to Readable's 12.9 and 15.5. Overall, the consistency with scores from readable.com validates the reliability of our approach.

\system's \textbf{entropy} measurements, ranging between 4.1-4.2, showed high consistency with Planetcalc's range of 4.1-4.3. Specifically, RedOrbit's privacy policy scored 4.1 with our method, versus 4.2 with Planetcalc. For Earthkam, Uptodate, and Amazon, both our method and Planetcalc reported an entropy score of 4.2, except for Amazon, where Planetcalc noted a slightly higher score of 4.3. This close alignment reinforces the accuracy and reliability of our method.

In evaluating ten policies, our method estimated a total \textbf{reading time} of 120 minutes and 5 seconds, closely paralleling The Read Time's calculation of 120 minutes and 40 seconds. The minor difference of 35 seconds between the two methods highlights the precision of our evaluation technique. The \textbf{unique words}, \textbf{keyword usage}, and \textbf{last update} features, which rely on searching specific keywords in the document, leverage a well-tested Python search library.


Finally, we evaluated the classifier's performance trained to predict a human analyst's \textbf{overall assessment} of 172 privacy policies. These policies correspond to devices analyzed by the Mozilla PNI initiative, along with their respective human analyst assessments. We employed cross-validation to ensure a reliable estimate of the model's effectiveness. The results, as shown in \Cref{table:performance_metrics}, indicate that \system accurately predicts human analyst assessments with a high F1-Score (0.91). Notably, \system shows better performance for the "acceptable" label compared to the "unacceptable" one. This discrepancy arises from the limited number of "unacceptable" instances in the dataset, leading to lower prediction accuracy due to model bias and challenges in identifying this minority class, despite implementing random oversampling.

\begin{figure}[!htb]
  \centering
  \includegraphics[width=1\linewidth,height=4cm,scale=2]{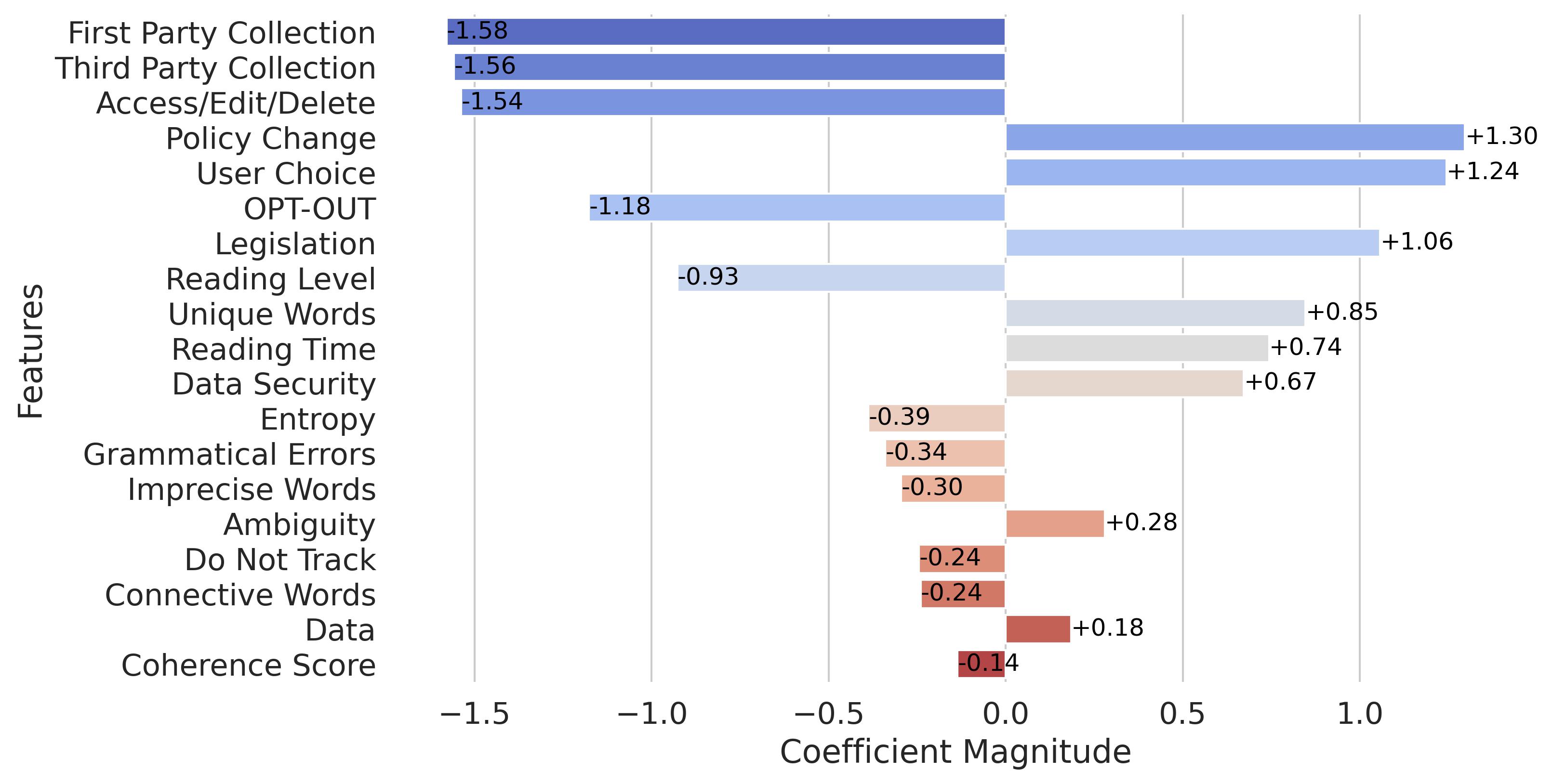}
  \caption{Feature Importance of the "overall assessment" model.}
  \label{fig:feat_imp}
\end{figure}

\Cref{fig:feat_imp} illustrates the significance of analyzing various features used in training the model. A positive feature importance coefficient implies an increase in the model's prediction accuracy with the feature value, while a negative coefficient suggests a decrease in accuracy, assuming other features remain constant. The findings reveal that features related to privacy analysis generally have a more substantial impact on prediction accuracy than those related to readability. However, features such as reading level, unique words, and reading time are also marked as highly important, hinting that an analyst's assessment may be influenced by the effort required to comprehend the policy. Additionally, it's noteworthy that PrivacyLens, mirroring the manual analysis of Mozilla PNI which took thousands of hours, demonstrates its efficiency in analyzing IoT device privacy policies.

\begin{table}[!htb]
\centering
\small
\begin{tabular}{l|c|c|c}
\hline
\textbf{Class} & \textbf{Precision} & \textbf{Recall} & \textbf{F1-score} \\ \hline
acceptable        & 0.96               & 0.96            & 0.96             \\
unacceptable        & 0.67               & 0.67            & 0.67             \\ \hline
Weighted Avg  &     0.92    &  0.90  &    0.91 \\
\hline
\end{tabular}
\caption{Performance of the "overall assessment" model.}
\label{table:performance_metrics}
\end{table}

\subsection{Feature Values Across Clusters}

\section{Summary of Policy Feature Analysis}

The analysis of policy features across different clusters revealed diverse characteristics in privacy policies. These are summarized based on the data presented in Tables \ref{tab:cluster1} to \ref{tab:cluster4}:

\begin{itemize}
    \item \textbf{Cluster 1 (Table \ref{tab:cluster1})}: Policies typically have moderate coherence scores (0.30-0.36), low frequency of imprecise (0.01-0.03) and connective words (0.03-0.05), and intermediate reading complexity (10.06-19.39). Reading times range from 3 to 8 minutes, with entropy values of 7.10-8.38 and unique word frequencies between 0.30 and 0.37. Grammatical errors vary from minimal to moderate (0.00-0.95).

    \item \textbf{Cluster 0 (Table \ref{tab:cluster0})}: This cluster features lower coherence scores (0.25-0.31), very low imprecise (0.00-0.02) and connective word frequencies (0.02-0.08), and higher reading complexities (12.10-26.99). Reading times are significantly longer (13-59 minutes), with high entropy (7.89-9.72) and unique word frequencies ranging from 0.14 to 0.34. Grammatical errors are minimal to moderate (0.00-0.92).

    \item \textbf{Cluster 2 (Table \ref{tab:cluster2})}: Policies have slightly higher coherence (0.26-0.34), imprecise word (0.01-0.03), and connective word frequencies (0.03-0.06) compared to Cluster 0. Reading complexities vary (7.66-19.19), with reading times ranging widely (7-48 minutes). Entropy values are 7.45-8.62, with unique word frequencies of 0.10-0.30. Grammatical errors range from none to high (0.00-0.98).

    \item \textbf{Cluster 3 (Table \ref{tab:cluster3})}: This cluster has the widest range in coherence scores (0.10-0.92), imprecise word (0.00-0.04), and connective word frequencies (0.00-0.09). Reading complexities are highly varied (4.70-86.68), as are reading times (0-39 minutes). Entropy values range from 5.23 to 9.71, and unique word frequencies vary significantly (0.10-0.77). Grammatical errors range from none to the maximum (0.00-1.00).

    \item \textbf{Cluster 4 (Table \ref{tab:cluster4})}: Policies in this cluster show moderate variability in coherence scores (0.25-0.64), imprecise word (0.00-0.03), and connective word frequencies (0.01-0.06). Reading complexities (9.34-27.44) and times (10-185 minutes) suggest varied policy lengths. Entropy is high (7.53-10.13), with unique word frequencies (0.07-0.29) and grammatical errors (0.00-0.96) being moderately varied.
\end{itemize}

\clearpage

\begin{figure}[!htb]
\centering

\begin{subfigure}{\textwidth}
\centering
\begin{tabular}{llll}
\toprule
Policy Features & Min Value & Median Value & Max Value \\
\midrule
Coherence Score & 0.30 & 0.35 & 0.36 \\
Freq. of Imprecise Words & 0.01 & 0.02 & 0.03 \\
Freq. of Connective Words & 0.03 & 0.04 & 0.05 \\
Reading Complexity & 10.06 & 14.44 & 19.39 \\
Reading Time (Min) & 3.00 & 4.00 & 8.00 \\
Entropy & 7.10 & 7.41 & 8.38 \\
Freq. of Unique Words & 0.30 & 0.33 & 0.37 \\
Grammatical Errors & 0.00 & 0.82 & 0.95 \\
\bottomrule
\end{tabular}
\caption{Cluster 1}
\label{tab:cluster1}
\end{subfigure}

\begin{subfigure}{\textwidth}
\centering
\begin{tabular}{llll}
\toprule
Policy Features & Min Value & Median Value & Max Value \\
\midrule
Coherence Score & 0.25 & 0.30 & 0.31 \\
Freq. of Imprecise Words & 0.00 & 0.01 & 0.02 \\
Freq. of Connective Words & 0.02 & 0.03 & 0.08 \\
Reading Complexity & 12.10 & 14.11 & 26.99 \\
Reading Time (Min) & 13.00 & 15.50 & 59.00 \\
Entropy & 7.89 & 8.12 & 9.72 \\
Freq. of Unique Words & 0.14 & 0.24 & 0.34 \\
Grammatical Errors & 0.00 & 0.82 & 0.92 \\
\bottomrule
\end{tabular}
\caption{Cluster 0}
\label{tab:cluster0}
\end{subfigure}

\begin{subfigure}{\textwidth}
\centering
\begin{tabular}{llll}
\toprule
Policy Features & Min Value & Median Value & Max Value \\
\midrule
Coherence Score & 0.26 & 0.31 & 0.34 \\
Freq. of Imprecise Words & 0.01 & 0.02 & 0.03 \\
Freq. of Connective Words & 0.03 & 0.04 & 0.06 \\
Reading Complexity & 7.66 & 12.65 & 19.19 \\
Reading Time (Min) & 7.00 & 17.00 & 48.00 \\
Entropy & 7.45 & 8.12 & 8.62 \\
Freq. of Unique Words & 0.10 & 0.22 & 0.30 \\
Grammatical Errors & 0.00 & 0.69 & 0.98 \\
\bottomrule
\end{tabular}
\caption{Cluster 2}
\label{tab:cluster2}
\end{subfigure}

\begin{subfigure}{\textwidth}
\centering
\begin{tabular}{llll}
\toprule
Policy Features & Min Value & Median Value & Max Value \\
\midrule
Coherence Score & 0.10 & 0.31 & 0.92 \\
Freq. of Imprecise Words & 0.00 & 0.02 & 0.04 \\
Freq. of Connective Words & 0.00 & 0.04 & 0.09 \\
Reading Complexity & 4.70 & 12.09 & 86.68 \\
Reading Time (Min) & 0.00 & 6.00 & 39.00 \\
Entropy & 5.23 & 7.72 & 9.71 \\
Freq. of Unique Words & 0.10 & 0.33 & 0.77 \\
Grammatical Errors & 0.00 & 0.50 & 1.00 \\
\bottomrule
\end{tabular}
\caption{Cluster 3}
\label{tab:cluster3}
\end{subfigure}

\begin{subfigure}{\textwidth}
\centering
\begin{tabular}{llll}
\toprule
Policy Features & Min Value & Median Value & Max Value \\
\midrule
Coherence Score & 0.25 & 0.31 & 0.64 \\
Freq. of Imprecise Words & 0.00 & 0.02 & 0.03 \\
Freq. of Connective Words & 0.01 & 0.04 & 0.06 \\
Reading Complexity & 9.34 & 13.24 & 27.44 \\
Reading Time (Min) & 10.00 & 21.00 & 185.00 \\
Entropy & 7.53 & 8.28 & 10.13 \\
Freq. of Unique Words & 0.07 & 0.20 & 0.29 \\
Grammatical Errors & 0.00 & 0.69 & 0.96 \\
\bottomrule
\end{tabular}
\caption{Cluster 4}
\end{subfigure}
\caption{Feature values across clusters}
\label{clusters_values}
\label{tab:cluster4}
\end{figure}
\clearpage

 \section{Country distribution}
\begin{figure*}[!h]
\centering
\includegraphics[width=1\linewidth,scale=1.5]{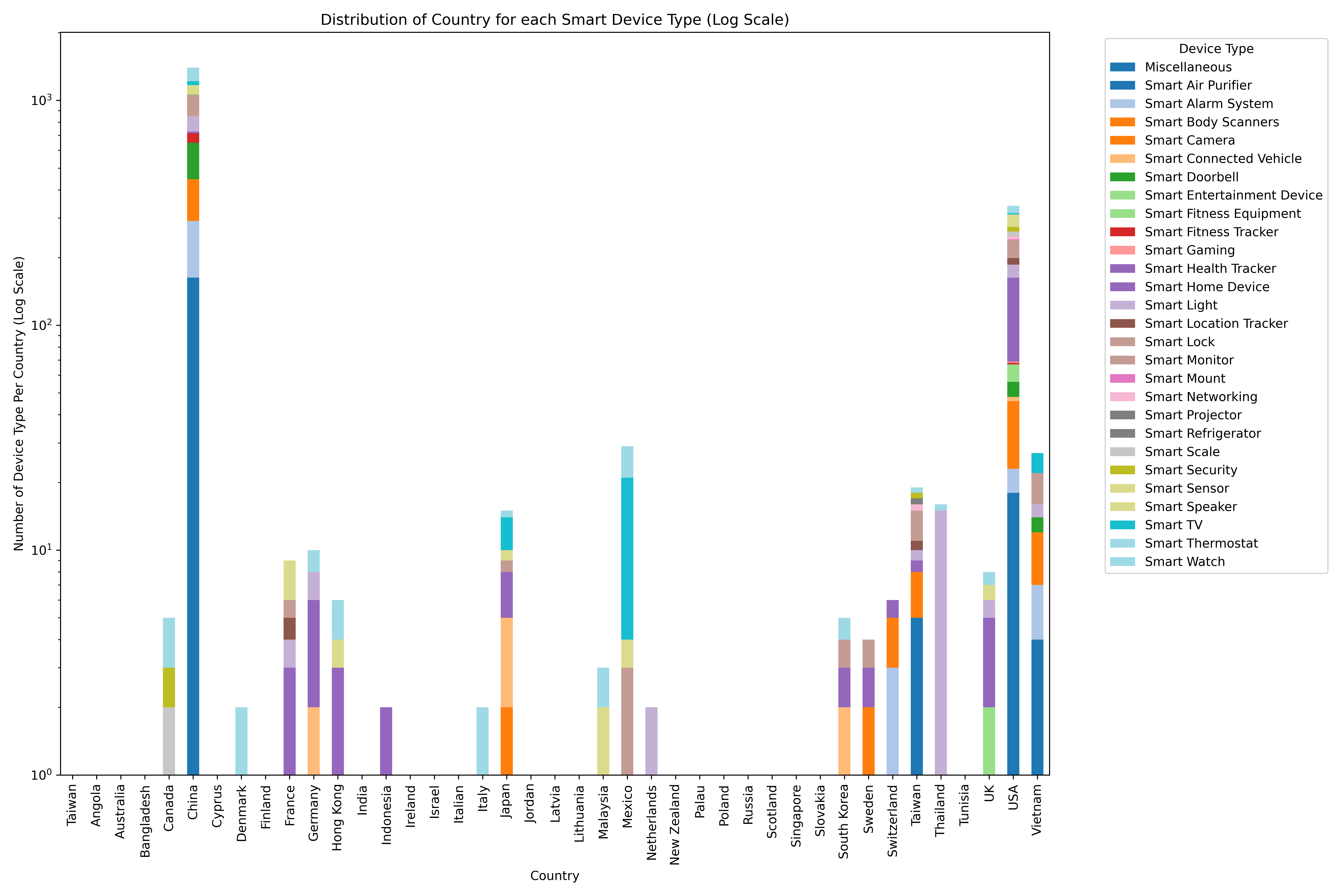}
 \caption{Distribution of Country for each Device Type}
 \label{fig:country_dev}
\end{figure*}

 Figure \ref{fig:country_dev} describes a comparative stacked bar chart of smart device type prevalence across multiple countries, plotted on a logarithmic scale to accommodate the wide variance in counts. Each bar represents a country, with color-coded segments indicating the count of each device type within that country. The chart highlights the heterogeneity in smart device distribution, reflecting varying degrees of market penetration and consumer adoption across the regions. This analysis provides valuable insights into the global distribution of smart devices and the differences in their adoption across different countries.

\section{Fakespot}
\begin{figure}[H]
    \centering
    \begin{tabular}{cc}
        \includegraphics[width=0.25\textwidth]{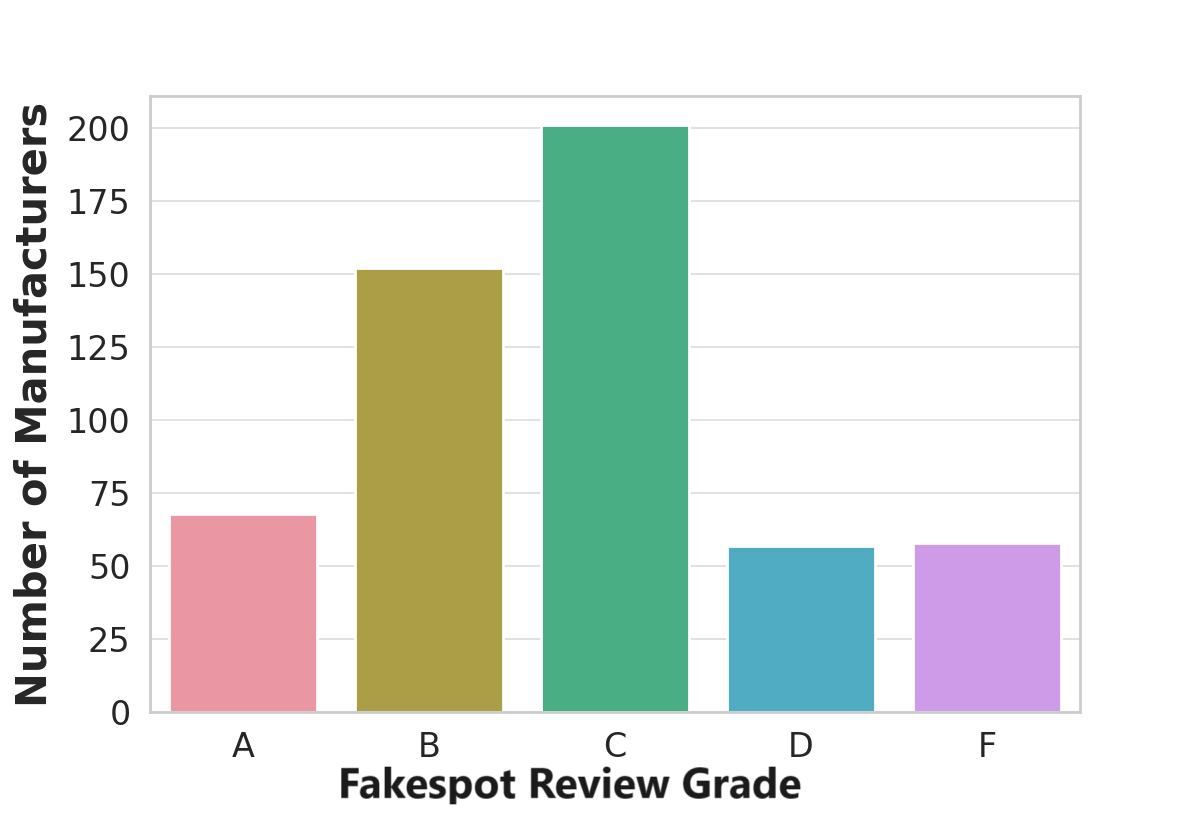} & \includegraphics[width=0.25\textwidth]{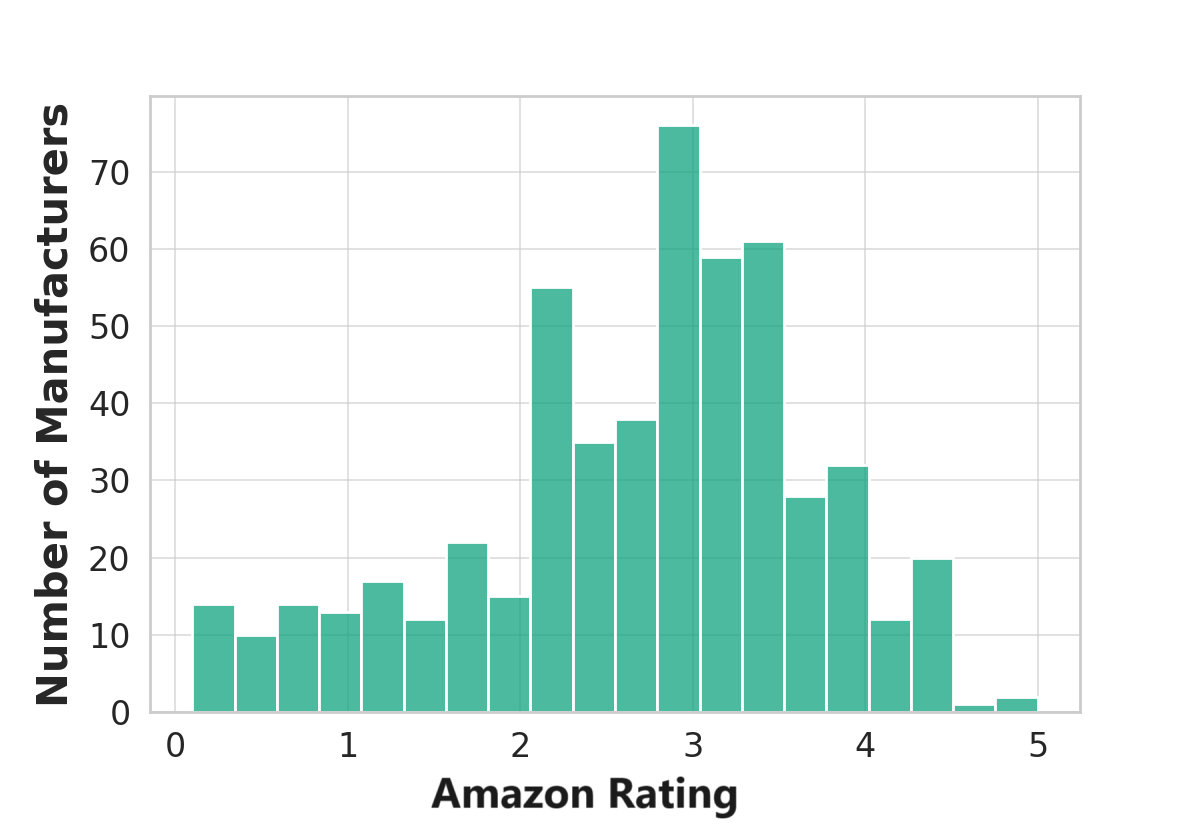} \\
        \includegraphics[width=0.25\textwidth]{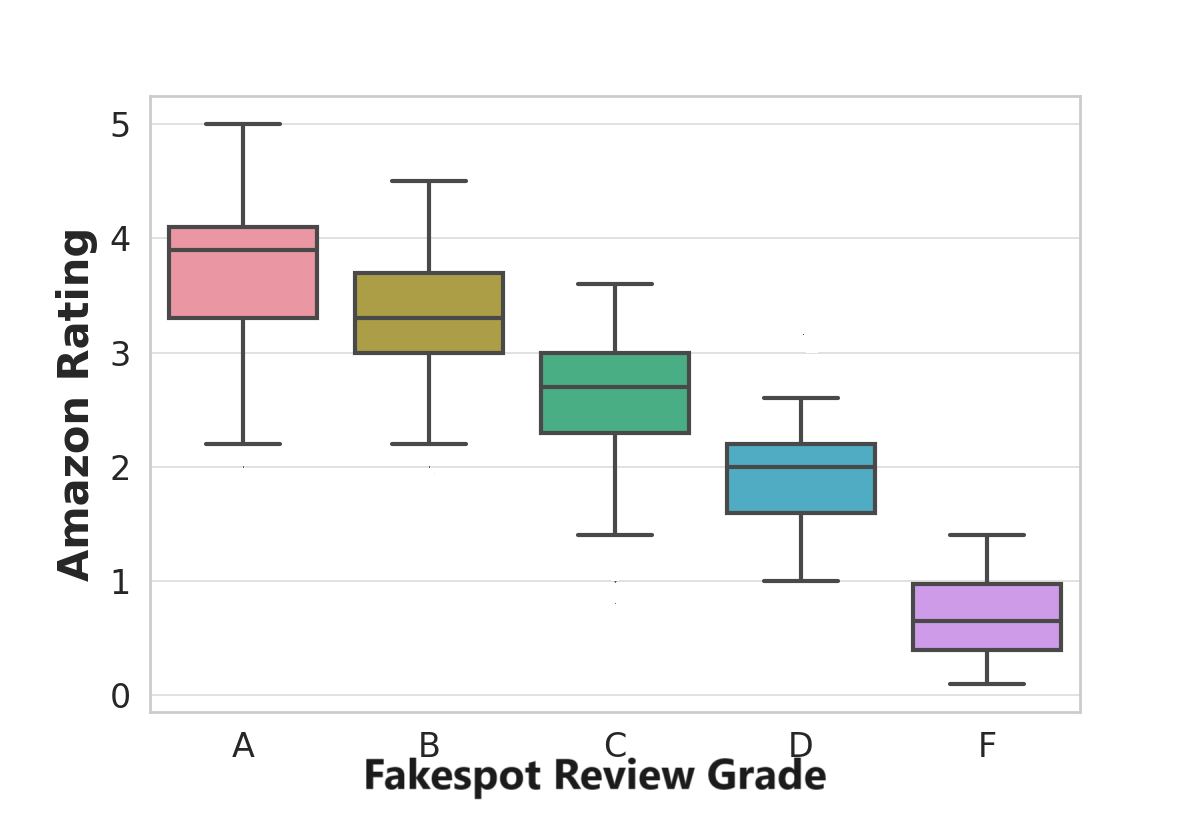} & \includegraphics[width=0.25\textwidth]{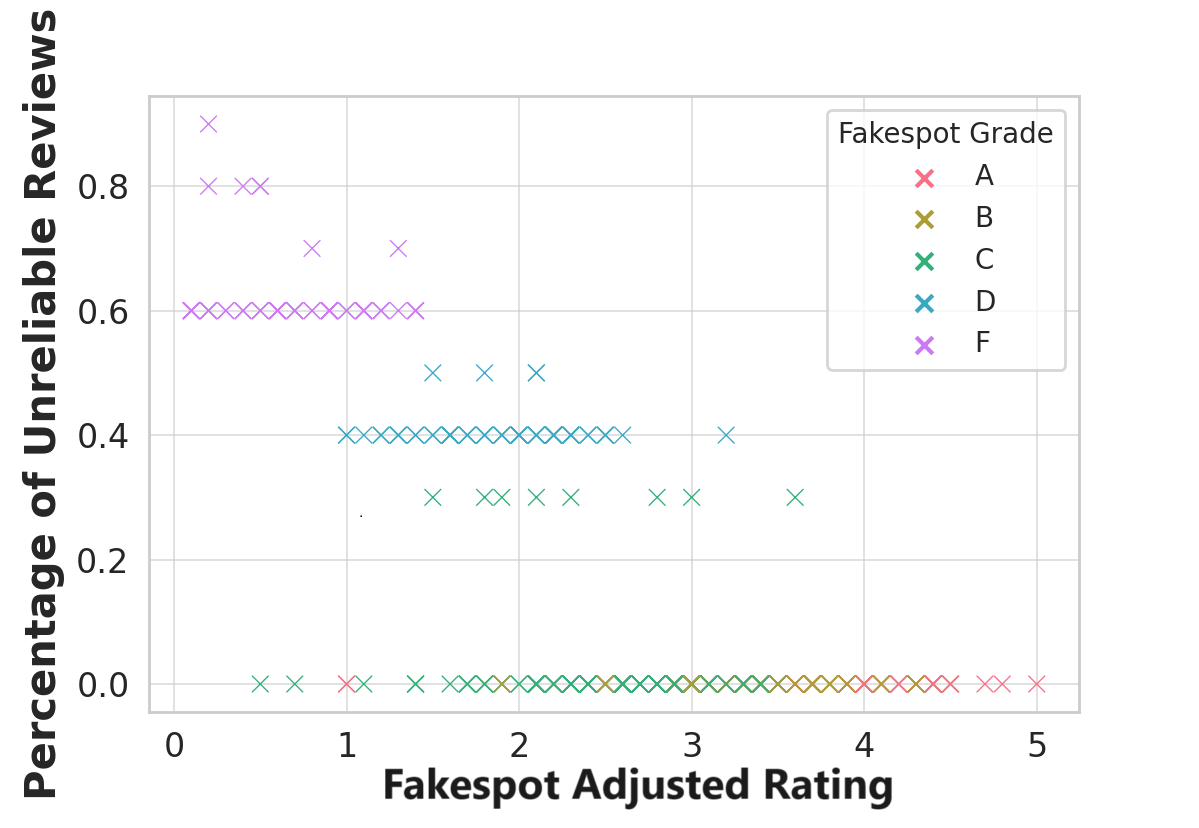}
    \end{tabular}
    \caption{Fakespot Company Review Grade Distribution.}
    \label{fig:fakespot_review}
\end{figure}

Charts in Figure \ref{fig:fakespot_review} provides a clear overview of the data collected from Fakespot. These figures illustrate various aspects of the analysis, such as the distribution of unreliable reviews, comparison of Amazon Ratings, and the overall Fakespot Company Review Grades for all 572 manufacturers that did not have a privacy policy.

\section{Policy Ambiguity Levels by Years}

Table\ref{tab:policy_ambiguity} provides the percentage of policy ambiguity levels for each year. The "Not Amb (\%)," "Somew. Amb (\%)," and "Very Amb (\%)" columns represent the percentage of policies that are not ambiguous, somewhat ambiguous, and very ambiguous, respectively. The "Total" column indicates the total number of policies analyzed for each year.
The data shows that the percentage of policies that are not ambiguous has fluctuated over the years, ranging from a low of 47.22\% in 2020 to a high of 71.43\% in 2018. The percentage of somewhat ambiguous policies has also varied, with a low of 5.56\% in 2017 and a high of 29.17\% in 2020. Similarly, the percentage of very ambiguous policies has ranged from 11.90\% in 2018 to 23.61\% in 2020.
This analysis provides valuable insights into the levels of policy ambiguity over the years, highlighting the fluctuating nature of policy clarity and the need for ongoing monitoring and analysis of policy implementation

\begin{table}[h]
\centering
\begin{tabular}{|c|c|c|c|c|}
\hline
Year & Not Amb (\%) & Somew. Amb (\%) & Very Amb (\%) & Total \\ \hline
2017 & 66.67 & 5.56  & 27.78 & 18 \\ 
2018 & 71.43 & 16.67 & 11.90 & 42 \\ 
2019 & 62.16 & 21.62 & 16.22 & 37 \\ 
2020 & 47.22 & 29.17 & 23.61 & 72 \\ 
2021 & 65.38 & 15.38 & 17.31 & 52 \\ 
2022 & 54.17 & 26.39 & 18.06 & 72 \\ 
2023 & 70.00 & 10.00 & 20.00 & 10 \\ \hline
\end{tabular}
\caption{Policy Ambiguity Levels by Year}
\label{tab:policy_ambiguity}
\end{table}

\begin{table}[h]
\centering
\begin{tabular}{lc}
\toprule
\textbf{IoT Keyword}    & \textbf{Weight} \\
\midrule
smart                    & 0.9             \\
wifi                     & 0.9             \\
bluetooth                & 0.9             \\
voice-controlled         & 0.9             \\
app-controlled           & 0.9             \\
chromecast               & 0.9             \\
hub                      & 0.7             \\
connected                & 0.7             \\
wireless                 & 0.65            \\
internet                 & 0.6             \\
remote                   & 0.65            \\
automation               & 0.7             \\
sensor                   & 0.7             \\
network                  & 0.65            \\
cloud                    & 0.6             \\
synchronization          & 0.6             \\
compatibility            & 0.55            \\
sync                     & 0.55            \\
interface                & 0.5             \\
dashboard                & 0.6             \\
real-time                & 0.6             \\
monitoring               & 0.65            \\
security                 & 0.65            \\
audio                    & 0.55            \\
microphone               & 0.55            \\
speaker                  & 0.55            \\
touch                    & 0.55            \\
gesture                  & 0.6             \\
light                    & 0.5             \\
api                      & 0.6             \\
sdk                      & 0.6             \\
protocol                 & 0.6             \\
ethernet                 & 0.6             \\
gateway                  & 0.7             \\
mesh                     & 0.7             \\
router                   & 0.6             \\
digital                  & 0.55            \\
interconnected           & 0.65            \\
\bottomrule
\end{tabular}
\caption{IoT Keywords Weights}
\label{tab:iot_keywords}
\end{table}

\begin{table}[h]
\centering
\begin{tabular}{l}
\toprule
\textbf{High Confidence Terms} \\
\midrule
iot                         \\
apple home kit              \\
alexa                       \\
google assistant            \\
android                     \\
\bottomrule
\end{tabular}
\caption{High Confidence Terms}
\label{tab:high_confidence_terms}
\end{table}

\begin{table}[h]
\centering
\begin{tabular}{lr}
\toprule
Category & Frequency \\
\midrule
Smart Speaker & 603 \\
Smart Thermostat & 579 \\
Smart Camera & 532 \\
Miscellaneous & 526 \\
Smart Lock & 447 \\
Smart Fitness Tracker & 430 \\
Smart Light & 366 \\
Smart Doorbell & 344 \\
Smart Alarm System & 343 \\
Smart TV & 275 \\
Smart Scale & 214 \\
Smart Home Device & 112 \\
Smart Air Purifier & 43 \\
Smart Sensor & 30 \\
Smart Watch & 30 \\
Smart Monitor & 24 \\
Smart Security & 21 \\
Smart Health Tracker & 19 \\
Smart Refrigerator & 19 \\
Smart Location Tracker & 16 \\
Smart Entertainment Device & 9 \\
Smart Connected Vehicle & 9 \\
Smart Networking & 8 \\
Smart Fitness Equipment & 3 \\
Smart Mount & 2 \\
Smart Projector & 2 \\
Smart Body Scanners & 1 \\
Smart Gaming & 1 \\
\bottomrule
\end{tabular}
\caption{Count of Smart Devices for Each Category}
\label{tab:cat_freq}
\end{table}

\section{Abiguity Analysis}
\Cref{tab:amb} shows the results in terms of ambiguity along with the accuracy of the classifiers used. For policies deemed 'Somewhat Ambiguous,' the F1 scores for the Random Forest Classifier and SVC improved to 0.89 and 0.94, respectively, an increase from the scores for 'Not Ambiguous' policies (0.78 and 0.85). However, this pattern shifted with 'Very Ambiguous' policies, where the F1 scores marginally decreased to 0.88 and 0.90. These findings suggest that while a moderate level of ambiguity may unexpectedly aid classification accuracy, high ambiguity poses significant challenges, indicating the need for more sophisticated algorithms capable of handling diverse ambiguity levels in policy texts

\begin{table}[!htb]
\centering
\resizebox{\linewidth}{!}{
\begin{tabular}{|c|c|c|c|}
\hline
\textbf{Ambiguity Class} & \textbf{Number of Policies} & \textbf{Random Forest Classifier} & \textbf{SVC} \\
\hline
Not Ambiguous & 471 & 0.78 & 0.85 \\
\hline
Somewhat Ambiguous & 186 & 0.89& 0.94 \\
\hline
Very Ambiguous & 162 & 0.88 & 0.90 \\
\hline
\end{tabular}
}
\caption{F1-score of ambiguity determination models.}
\label{tab:amb}
\end{table}

\section{Keyword Usage Across Clusters}

The analysis of keyword usage across different clusters revealed diverse characteristics in privacy policies. These are summarized based on the data presented in Tables \ref{tab:l1} to \ref{tab:l5}:

\begin{itemize}
    \item \textbf{Cluster 1 (Table \ref{tab:l1})}: This cluster shows moderate to high usage across all keyword categories, with the highest average usage in the "Collection," "Purpose," and "Retention" keyword categories. This suggests a consistent and relatively high level of usage across a wide range of keywords in this cluster.

    \item \textbf{Cluster 0 (Table \ref{tab:l2})}: In this cluster, the usage of all keyword categories is generally lower compared to Cluster 1. The "Collection" keyword category still has the highest average usage, but the usage of other categories is notably lower. This cluster represents a lower overall usage of keywords across all categories.

    \item \textbf{Cluster 2 (Table \ref{tab:l3})}: Cluster 3 exhibits the highest usage across all keyword categories, with the "Collection," "Sharing," "Purpose," and "Access" categories showing particularly high average usage. This cluster represents a high overall usage of keywords, especially in the specified categories.

    \item \textbf{Cluster 3 (Table \ref{tab:l4})}: This cluster shows low to moderate usage across all keyword categories, with the "Collection" category having the highest average usage. The usage of other categories is relatively low. This cluster represents a lower overall usage of keywords, similar to Cluster 2.

    \item \textbf{Cluster 4 (Table \ref{tab:l5})}:  Cluster 5 demonstrates the highest usage across all keyword categories, with the "Collection," "Sharing," and "Purpose" categories showing particularly high average usage. This cluster represents the highest overall usage of keywords, especially in the specified categories.
\end{itemize}

In summary, the cluster analysis reveals distinct patterns of keyword usage across different clusters, with some clusters showing higher overall usage of keywords across all categories, while others exhibit lower usage. These findings provide valuable insights into the distribution and usage patterns of the specified keyword categories within the dataset.

\begin{table}[h]
\centering
\caption{Cluster 0}
\begin{tabular}{|l|l|l|l|}
\hline
\textbf{Keyword Categories} & \textbf{Min Usage} & \textbf{Median Usage} & \textbf{Max Usage} \\ \hline
Collection Keywords & 44.00 & 46.00 & 186.00 \\
Sharing Keywords & 42.00 & 51.00 & 222.00 \\
Purpose Keywords & 53.00 & 64.50 & 158.00 \\
Access Keywords & 45.00 & 81.00 & 114.00 \\
Security Keywords & 16.00 & 16.50 & 64.00 \\
Policy Change Keywords & 23.00 & 41.00 & 90.00 \\
Legislation Keywords & 0.00 & 0.00 & 0.00 \\
Choice Keywords & 22.00 & 32.50 & 169.00 \\
Retention Keywords & 9.00 & 12.50 & 104.00 \\
IoT Data Keywords & 0.00 & 0.00 & 9.00 \\ \hline
\end{tabular}
\label{tab:l1}
\end{table}

\begin{table}[h]
\centering
\caption{Cluster 1}
\begin{tabular}{|l|l|l|l|}
\hline
\textbf{Keyword Categories} & \textbf{Min Usage} & \textbf{Median Usage} & \textbf{Max Usage} \\ \hline
Collection Keywords & 34.00 & 45.00 & 65.00 \\
Sharing Keywords & 6.00 & 14.00 & 25.00 \\
Purpose Keywords & 13.00 & 33.00 & 46.00 \\
Access Keywords & 8.00 & 11.00 & 20.00 \\
Security Keywords & 3.00 & 4.00 & 11.00 \\
Policy Change Keywords & 6.00 & 8.00 & 17.00 \\
Legislation Keywords & 0.00 & 0.00 & 0.00 \\
Choice Keywords & 4.00 & 11.00 & 23.00 \\
Retention Keywords & 4.00 & 10.00 & 16.00 \\
IoT Data Keywords & 0.00 & 0.00 & 4.00 \\ \hline
\end{tabular}
\label{tab:l2}
\end{table}

\begin{table}[h]
\centering
\caption{Cluster 2}
\begin{tabular}{|l|l|l|l|}
\hline
\textbf{Keyword Categories} & \textbf{Min Usage} & \textbf{Median Usage} & \textbf{Max Usage} \\ \hline
Collection Keywords & 63.00 & 143.00 & 385.00 \\
Sharing Keywords & 22.00 & 74.50 & 225.00 \\
Purpose Keywords & 24.00 & 90.00 & 245.00 \\
Access Keywords & 22.00 & 53.00 & 210.00 \\
Security Keywords & 6.00 & 38.50 & 110.00 \\
Policy Change Keywords & 8.00 & 29.00 & 111.00 \\
Legislation Keywords & 0.00 & 0.00 & 0.00 \\
Choice Keywords & 3.00 & 33.50 & 125.00 \\
Retention Keywords & 50.00 & 108.50 & 447.00 \\
IoT Data Keywords & 0.00 & 0.00 & 21.00 \\\hline
\end{tabular}
\label{tab:l3}
\end{table}

\begin{table}[h]
\centering
\caption{Cluster 3}
\begin{tabular}{|l|l|l|l|}
\hline
\textbf{Keyword Categories} & \textbf{Min Usage} & \textbf{Median Usage} & \textbf{Max Usage} \\ \hline
Collection Keywords & 0.00 & 50.00 & 138.00 \\
Sharing Keywords & 0.00 & 24.00 & 103.00 \\
Purpose Keywords & 0.00 & 32.00 & 175.00 \\
Access Keywords & 0.00 & 18.00 & 82.00 \\
Security Keywords & 0.00 & 13.00 & 77.00 \\
Policy Change Keywords & 0.00 & 14.00 & 141.00 \\
Legislation Keywords & 0.00 & 0.00 & 0.00 \\
Choice Keywords & 0.00 & 9.00 & 82.00 \\
Retention Keywords & 0.00 & 12.00 & 63.00 \\
IoT Data Keywords & 0.00 & 0.00 & 12.00 \\ \hline
\end{tabular}
\label{tab:l4}
\end{table}

\begin{table}[h]
\centering
\caption{Cluster 4}
\begin{tabular}{|l|l|l|l|}
\hline
\textbf{Keyword Categories} & \textbf{Min Usage} & \textbf{Median Usage} & \textbf{Max Usage} \\ \hline
Collection Keywords & 83.00 & 201.00 & 1,145.00 \\
Sharing Keywords & 29.00 & 110.00 & 815.00 \\
Purpose Keywords & 49.00 & 134.00 & 786.00 \\
Access Keywords & 27.00 & 84.00 & 673.00 \\
Security Keywords & 15.00 & 42.00 & 254.00 \\
Policy Change Keywords & 11.00 & 46.00 & 310.00 \\
Legislation Keywords & 0.00 & 0.00 & 0.00 \\
Choice Keywords & 12.00 & 45.00 & 375.00 \\
Retention Keywords & 8.00 & 52.00 & 758.00 \\
IoT Data Keywords & 0.00 & 1.00 & 58.00 \\ \hline
\end{tabular}
\label{tab:l5}
\end{table}

\end{appendices}

\end{document}